\documentclass[sigconf]{acmart}

\setcopyright{none}

\usepackage{graphicx}
\usepackage{enumerate}

\usepackage{amssymb}
\usepackage{amsmath}
\usepackage{bbm}
\usepackage{float}
\usepackage{verbatim}
\usepackage{xspace,afterpage}
\usepackage{ragged2e}
\usepackage{tablefootnote}
\usepackage{paralist}

\usepackage{subcaption}

\usepackage{dsfont}

\usepackage[ruled,vlined]{algorithm2e}
\usepackage{amsthm} 

\usepackage{soul}
            
\newtheorem{theo}{Theorem}

\newtheorem{prop}{Proposition}

\newtheorem{cor}{Corollary}

\newtheorem{lem}{Lemma}

\AtBeginDocument{%
  \providecommand\BibTeX{{%
    \normalfont B\kern-0.5em{\scshape i\kern-0.25em b}\kern-0.8em\TeX}}}

\setcopyright{none}
\copyrightyear{none}
\acmYear{none}
\acmDOI{none}

\acmSubmissionID{\#34}

\begin{document}
\title{
Congestion-aware routing and content placement\\ in elastic cache networks}

\author{Jinkun Zhang}
\affiliation{%
  \institution{Northeastern University}
  \country{}}
\email{jinkunzhang@ece.neu.edu}

\author{Edmund Yeh}
\affiliation{%
  \institution{Northeastern University}
  \country{}}
 \email{eyeh@ece.neu.edu}


\begin{abstract}
Caching can be leveraged to significantly improve network performance and mitigate congestion. 
However, characterizing the optimal tradeoff between routing cost and cache deployment cost remains an open problem.
In this paper, for a network with arbitrary topology and congestion-dependent nonlinear cost functions, we aim to jointly determine the cache deployment, content placement, and hop-by-hop routing strategies, so that the sum of routing cost and cache deployment cost is minimized.
{ We tackle this NP-hard problem starting with a fixed-routing setting, and then to a general dynamic-routing setting. 
For the fixed-routing setting, a Gradient-combining Frank-Wolfe algorithm with $(\frac{1}{2},1)$-approximation is presented.
For the general dynamic-routing setting, we obtain a set of KKT necessary optimal conditions, and devise a distributed and adaptive online algorithm based on the conditions.}
We demonstrate via extensive simulation that our algorithms significantly outperform a number of baseline techniques.
\end{abstract}



\maketitle
\section{Introduction}
With the explosive growth of Internet traffic volume, caching, by bringing popular content closer to consumers, is recognized as one of the most efficient ways to mitigate bandwidth bottlenecks and reduce delay in modern content delivery networks.
However, as a resource, network caches have neither prescribed sizes, nor are they provided for free.
The network operator may pay a cost to deploy (e.g., rent from service providers or purchase and install manually) caches of elastic sizes across the network, if this yields satisfactory improvement in network performance (e.g., average delay).
Therefore, a rational network operator may seek to quantify the tradeoff between cache deployment costs and network performance metrics. 
In this paper, by formulating and solving the problem of joint routing and caching with elastic cache sizes, we help answer the question:
\emph{is it worth deploying more cache}?

On the one hand, with fixed cache capacities, joint optimization of routing and caching is extensively studied in various real-life networking contexts, such as content delivery networks (CDNs) \cite{dehghan2015complexity} and information-centric networks (ICNs) \cite{zhang2014named}. 
Routing cost, e.g., average packet delay, is one of the most important network performance metrics, and is frequently selected as optimization objective. 
On the other hand, optimization over elastic cache sizes has drawn significant attention recently, meeting the demand of rapidly growing small content providers tending to lease storage from elastic CDNs (e.g., Akamai Aura) instead of purchasing and maintaining by themselves.
Tradeoff between cache deployment cost and cache utilities is studied for simple topologies \cite{ye2021joint,dehghan2019utility}.

However, higher cache utility (more cache hits or higher cache hit ratio) does not always imply lower routing costs. 
For example, requests served with a hit ratio $1$ at distant servers could incur higher routing costs than requests served locally with a lower hit ratio.
It is more directly in the network operator's interest to achieve lower routing costs, 
e.g., lower user latency or link usage fee.
To our knowledge, the tradeoff between routing cost and cache deployment cost remains an open problem.
In this paper, 
we fill this gap by minimizing a total cost -- the sum of network routing cost and cache deployment cost, in networks with arbitrary topology and general convex cost functions.


We consider a cache-enabled content delivery network with arbitrary multi-hop topology and stochastic, stationary request arrivals.
Each request is routed in a hop-by-hop manner until it reaches 
either a node that caches the requested content, or a designated server that permanently keeps the content.
The content is then sent back to the requester along the reverse path.
Convex congestion-dependent costs are incurred on the links due to transmission, and at the nodes due to cache deployment.
Our objective is to devise a distributed and adaptive online algorithm determining the routing and caching strategies, so that the total cost is minimized.  

We study the proposed problem first in a fixed-routing setting and then in a dynamic-routing setting.
For the fixed-routing case, we achieve a $(\frac{1}{2},1)$ approximation by using the Gradient-combining Frank-Wolfe algorithm proposed in \cite{mitra2021submodular}.
The general dynamic-routing setting can be reduced to the congestion-dependent joint routing and caching problem in \cite{mahdian2018mindelay} if the cache sizes are fixed.
This problem has no known solution with a constant factor approximation.
Nevertheless, inspired by \cite{gallager1977minimum},  we propose a method which differs from \cite{mahdian2018mindelay} and provides stronger theoretical insight.

Specifically, we propose a modification to the KKT necessary condition for the general dynamic-routing setting.
The modified condition is a more restrictive version of KKT condition, which avoids particular saddle points.
It suggests each node handles arrival requests in the way that achieves minimum marginal cost -- either by forwarding to a nearby node or by expanding the local cache.
We show that the total cost lies within a finite bound from the global optimum if the modified condition is satisfied, and the bound meets $0$ in some special cases.

Moreover, a distributed online algorithm can be developed based on the modified condition. 
The algorithm allows nodes to dynamically adjust their routing and caching strategies, adapting to moderate changes in request rates and cost functions.



The main contributions of this paper are as follows:
\begin{itemize}
    \item We propose a mathematical framework unifying the cache deployment, content placement and routing strategies in a network with arbitrary topology and general convex costs.
    We then propose the total cost minimization problem, which is shown to be NP-hard.
    
    \item We first study the fixed-routing setting.
    We recast the proposed problem into a \emph{DR-submodular + concave} maximization, then provide a Gradient-combining Frank-Wolfe algorithm with $(\frac{1}{2},1)$ approximation.
    
    \item For the general case, we develop a modification to the KKT necessary condition. 
    
    
    \item We propose a distributed and adaptive online gradient projection algorithm that converges to the modified condition, with novel loop-prevention and rounding mechanisms.
    
    \item With a packet-level simulator, we compare proposed algorithms against baselines in multiple scenarios. The proposed algorithms show significant performance improvements.
\end{itemize}

The remainder of this paper is organized as follows.
In Section \ref{sec:related work} we give a brief review of related works. 
In Section \ref{sec:model} we present our model and formulate the problem.
We study the fixed-routing case in Section \ref{sec:fixed routing}.
For the general case, we propose the KKT conditions in Section \ref{sec:general condition}, and develop the online algorithm in Section \ref{Sec:GeneralAlgorithm}.
We present our simulation results in Section \ref{sec:simulation},
discuss potential extentions in Section \ref{sec:extensions},
and conclude the paper in Section \ref{sec:conclusion}.

\section{Related works }
\label{sec:related work}

\noindent\textbf{Routing in cache-enabled networks.}
Routing and caching strategies are often managed separately in practical usage, for example, traditional priority-based cache replacement policies (e.g.,  First In First Out (FIFO), Least Recently Used (LRU) 
, Least Frequently Used (LFU) 
), combined with shortest path routing and its extensions.
Gallager \cite{gallager1977minimum} provided the global optimal solution to the multi-commodity routing problem with arbitrary network topology and general convex costs using a distributed hop-by-hop algorithm.
The caching problem with fixed routing path is shown NP-complete \cite{shanmugam2013femtocaching} even with linear link costs, and a distributed online algorithm \cite{ioannidis2018adaptive} achieves a $1-1/e$ approximation.

Nevertheless, jointly-designed routing and caching strategies can reduce routing costs significantly compared to those designed separately.
Existing joint strategies show enormous diversity in network topology, objective metric, and mathematical technique. 
A throughput-optimal dynamic forwarding and caching algorithm for ICN is proposed by  \cite{yeh2014vip}. 
Ioannidis and Yeh \cite{ioannidis2017jointly} extended \cite{ioannidis2018adaptive} to joint routing and routing problem with linear link costs.
Nevertheless, the routing cost-optimal joint routing and caching with arbitrary topology and convex costs remains an open problem.
Mahdian and Yeh \cite{mahdian2018mindelay} first formulated this problem in a hop-by-hop manner and devised a heuristic algorithm, however, without an analytical performance guarantee.

\noindent\textbf{Elastic cache sizes.}
A number of works on elastic caching recently emerged.
One line focused on jointly optimizing cache deployment and content placement subject to a total budget constraint. 
Mai et al. \cite{mai2019optimal} extended the idea in \cite{ioannidis2018adaptive} with a game-theory based framework. 
Kwak et al. \cite{kwak2021elastic} generalized FemtoCaching to elastic sizes.
Yao et al. \cite{yao2018joint} and Dai et al. \cite{dai2018optimized} studied content placement and storage allocation in cloud radio access network (C-RAN). 
Peng et al. \cite{peng2016cache} and Liu et al. \cite{liu2016much} studied cache allocation in backhaul-limited wireless networks.

Another line considered the tradeoff between cache deployment cost and cache utilities, so the total budget itself is optimized to pursue a maximum gain.
Chu et al. \cite{chu2018joint} maximized linear utilities of content providers over cache size and content placement, and is extended by Dehghan et al. \cite{dehghan2019utility} to concave utilities.
Ma et al. \cite{ma2015cashing} developed a cache monetizing scheme.
Recently, Ye et al. \cite{ye2021joint} optimized cache size scaling and content placement via learning.
In this line, almost all cache utilities are defined as functions of cache hit count or ratio, whereas, in a network of arbitrary topology and convex congestion-dependent link costs, higher cache hit count or ratio does not necessarily yield lower routing cost.

This paper differs from previous works as we simultaneously
\begin{inparaenum}[(i)]
\item assume an arbitrary multi-hop network topology,
\item adopt a hop-by-hop routing scheme with congestion-dependent costs,
\item consider elastic cache sizes with convex deployment costs,
\item achieve tradeoff between network performance and cache deployment costs instead of operating with a fixed budget, and
\item incorporate routing cost as the performance metric instead of cache utilities.
\end{inparaenum}

\begin{figure}[h]
  \centering
  \includegraphics[width=1\linewidth]{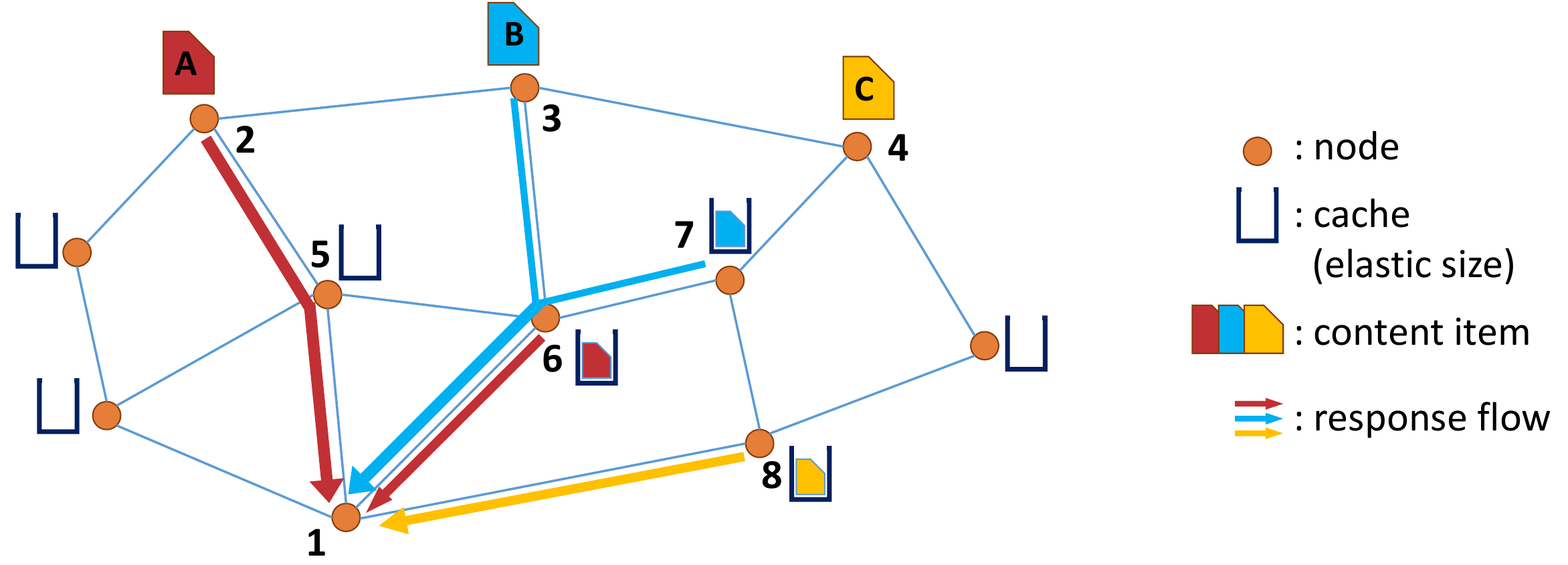}
  \caption{An example network. Node $1$ makes requests for item $A$, $B$ and $C$, designated servers are at node $2$, $3$ and $4$, respectively. Node $6$, $7$, $8$ cache the items. 
  Node $1$ may forward a fraction of requests for $A$ to $5$ as link $(6,1)$ is congested.}
 \label{fig_sample_network}
\end{figure}

\section{Model and problem formulation}
\label{sec:model}


\subsection{Cache-enabled network}
We model a cache-enabled network by a directed graph $\mathcal{G} = (\mathcal{V},\mathcal{E})$, where $\mathcal{V}$ is the set of nodes and $\mathcal{E}$ is the set of directed links. 
We assume $(j,i) \in \mathcal{E}$ for any $(i,j) \in \mathcal{E}$.
For node $i \in \mathcal{V}$, let $\mathcal{N}(i) = \left\{j \in \mathcal{V} \big| (i,j) \in \mathcal{E}\right\} = \left\{j \in \mathcal{V} \big| (j,i) \in \mathcal{E}\right\}$ denote the neighbors of $i$.

Let $\mathcal{C}$ denote the content items, i.e., the \emph{catalog}. 
We assume all items are of equal size $L_{\text{item}} = 1$.\footnote{Contents of non-equal sizes can be partitioned into chunks of equal size.}
Items are permanently kept at their \emph{designated servers}, without consuming the servers' cache space.
Let set $\mathcal{S}_k \subseteq \mathcal{V}$ be the designated server(s) for item $k \in \mathcal{C}$.

Nodes have access to caches of elastic size, and can optionally store content items by consuming corresponding cache space. 
We denote node $i$'s \emph{cache decisions} by $\boldsymbol{x}_i = [x_{i}(k)]_{k \in \mathcal{C}}$, where the binary  decision $x_{i}(k) \in \{0,1\}$ indicates whether node $i$ choose to cache item $k$ (i.e., $x_{i}(k)=1$ if node $i$ caches item $k$). 
We denote by $\boldsymbol{x} = [x_{i}(k)]_{i\in\mathcal{V},k \in \mathcal{C}}$ the \emph{global caching decision}.

\subsection{Request and response routing}
Packet transmission in $\mathcal{G}$ is request driven. 
We use $(i,k)$ to denote the request made by node $i$ for item $k$, 
and assume that request packets of $(i,k)$ is generated by $i$ at a steady \emph{exogenous request input rate} $r_{i}(k)$ (request packet/sec).
Request packets are routed in $\mathcal{G}$ in a hop-by-hop manner. 
Let $t_i(k)$ be the total request arrival rate for item $k$ at node $i$. That is, $t_i(k)$ includes node $i$'s exogenous request input rate $r_i(k)$, and the rate of endogenously arrival requests forwarded from other nodes to node $i$.
Of the request packets of item $k$ that arrive at node $i$, 
a fraction of $\phi_{ij}(k) \in [0,1]$ is forwarded to neighbor $j \in \mathcal{N}(i)$.
Thus for any $i \in \mathcal{V}$ and $k\in\mathcal{C}$,
\begin{equation*}
    t_i(k) = r_i(k) + \sum\nolimits_{j \in \mathcal{V}}t_j(k)\phi_{ji}(k),
\end{equation*}
where $\phi_{ij}(k) \equiv 0$ if $(i,j) \not\in \mathcal{E}$.
We denote $i$'s \emph{routing strategy} by $\boldsymbol{\phi}_i = [\phi_{ij}(k)]_{j \in \mathcal{V}, k \in \mathcal{C}}$, and denote the \emph{global routing strategy} by $\boldsymbol{\phi} = [\phi_{ij}(k)]_{i,j\in\mathcal{V},k\in\mathcal{C}}$.
Request packets for item $k$ terminate at node $i$ if $i$ caches $k$ or $i$ is a designated server of $k$. 
Thus the flow-conservation holds for all $i \in \mathcal{V}$ and $k \in \mathcal{C}$, 
\begin{equation}
\begin{aligned}
    x_{i}(k) + \sum\nolimits_{j \in \mathcal{V}} \phi_{ij}(k) = \begin{cases}
    1, \quad \text{ if } i \not\in \mathcal{S}_k,
    \\ 0, \quad \text{ if } i \in \mathcal{S}_k.
    \end{cases}
\end{aligned}
    \label{FlowConservation_cache_origin}
\end{equation}

Suppose node $i$ is not a designated server of item $k$, then constraint \eqref{FlowConservation_cache_origin} implies that: 
if $i$ does not cache $k$, every request packet for $k$ arriving at $i$ must be forwarded to one of the neighbors; if $i$ caches $k$, all request packets for $k$ arriving at $i$ terminate there.


When a request packet terminates, a response packet is generated and delivers the requested item back to requester in the reverse path.\footnote{
Such mechanism is implemented in ICN with the Forwarding Interest Base (FIB) and Pending Interest Table (PIT) \cite{yeh2014vip}.
We do not consider request aggregation in this paper. Different request packets for the same item are recorded and routed separately.
}
An example network is shown in Fig. \ref{fig_sample_network}.

\subsection{Routing and cache deployment costs}
Costs are incurred on the links due to packet transmission or packet queueing, and at the nodes due to cache deployment.
Since the size of request packets is typically negligible compared to the size of responses carrying content items, we consider only the link cost caused by responses.

Let $f_{ij}(k)$ be the rate of responses (response packet/sec) traveling through link $(i,j)$ carrying item $k$.
Recall that $L_{\text{item}}=1$, then $f_{ij}(k)$ equals the flow rate on link $(i,j)$ due to item $k$.
Let $F_{ij}$ be the total flow rate on $(i,j)$.
Since each request packet forwarded from $i$ to $j$ must fetch a response packet travelling through $(j,i)$, we have
\begin{equation*}
        f_{ji}(k) = t_i(k) \phi_{ij}(k), \quad F_{ij} = \sum\nolimits_{k \in \mathcal{C}} f_{ij}(k).
\end{equation*}

We denote by $D_{ij}(F_{ij})$ the routing cost on link $(i,j)$, and assume the cost function $D_{ij}(\cdot)$ is continuously differentiable, monotonically increasing and convex, with $D_{ij}(0) = 0$.
Such $D_{ij}(\cdot)$ subsumes a variety of existing cost functions, including commonly adopted linear cost 
or transmission latency \cite{xiang2020joint}. 
It can also approximate the link capacity constraint $F_{ij} \leq C_{ij}$ (e.g., in \cite{liu2019joint}) by selecting a smooth convex function that goes to infinity as $F_{ij}$ approaches $C_{ij}$. 
It also incorporates congestion-dependent performance metrics.
For example, let $\mu_{ij}$ be the service rate of an M/M/1 queue, $D_{ij}(F_{ij}) = {F_{ij}}/\left({\mu_{ij}-F_{ij}}\right)$ gives the average number of packets waiting for or being served in the queue \cite{bertsekas2021data}, and the aggregated cost $\sum_{(i,j) \in \mathcal{E}}{F_{ij}}/\left({\mu_{ij}-F_{ij}}\right)$, by Little's Law, is proportional to the expected system latency of packets in the network.

On the other hand, the cache occupancy at node $i$ is given by 
\begin{equation*}
    X_i = \sum\nolimits_{k \in \mathcal{C}}x_i(k).
\end{equation*}

We denote by $B_{i}(X_{i})$ the cache deployment cost at node $i$, and also assume $B_{i}(\cdot)$ to be continuously differentiable, monotonically increasing and convex, with $B_{i}(0) = 0$.
Cache deployment cost can represent the money expense to buy/rent storage (e.g., \cite{chu2018joint,ye2021joint,dehghan2019utility}), 
or approximate traditional hard cache capacity constraints.


\subsection{Problem formulation}
We aim to jointly optimize cache decisions and routing strategies to minimize the total cost. Nevertheless, to make progress toward an approximate solution of the mixed-integer non-linear problem, we relax binary $x_{i}(k)$ into the continuous $y_{i}(k) \in [0,1]$.
We denote node $i$'s \emph{caching strategy} by $\boldsymbol{y}_i = [y_{i}(k)]_{k \in \mathcal{C}}$, and the \emph{global caching strategy} by $\boldsymbol{y} = [y_{i}(k)]_{i \in \mathcal{V}, k \in \mathcal{C}}$.
Such continuous relaxation is widely adopted, e.g. \cite{ioannidis2018adaptive,liu2019joint}, and can be practically realized by a probabilistic caching scheme, i.e., node $i$ independently caches item $k$ with probability $y_{i}(k) = \mathbb{E}[x_{i}(k)]$.
We discuss other rounding techniques and provide a distributed randomized rounding algorithm in Section \ref{section:rounding}.

Let $Y_i = \sum_{k \in \mathcal{C}}y_i(k)$, and rewrite \eqref{FlowConservation_cache_origin} as follows,
\begin{equation}
\begin{aligned}
        y_{i}(k) + \sum\nolimits_{j \in \mathcal{V}} \phi_{ij}(k) = 
        \begin{cases}
        1, \quad \text{if } i \not\in \mathcal{S}_k,
\\        0, \quad \text{if } i \in \mathcal{S}_k.
        \end{cases}
\end{aligned}
    \label{FlowConservation_cache}
\end{equation}
The joint routing and content placement problem is formulated as
\begin{equation}
    \begin{aligned}
        \min_{\boldsymbol{\phi},\boldsymbol{y}} \quad &T(\boldsymbol{\phi},\boldsymbol{y}) = \sum\nolimits_{(i,j) \in \mathcal{E}} D_{ij}(F_{ij}) + \sum\nolimits_{i\in\mathcal{V}}B_{i}(Y_{i})
        \\\text{subject to} \quad 
        & 0 \leq \phi_{ij}(k) \leq 1, \quad \forall (i,j) \in \mathcal{E}, k \in \mathcal{C} 
        \\ & 0 \leq y_{i}(k) \leq 1, \quad \forall i \in \mathcal{V}, k \in \mathcal{C}
        \\ & \text{ \eqref{FlowConservation_cache} holds.}
    \end{aligned}
    \label{Objective_cache}
\end{equation}

\begin{prop}
Problem \eqref{Objective_cache} is NP-hard.
\label{Prop_cache_NPhard}
\end{prop}
The proof is provided in Appendix \ref{prof_prop_NPhard}.
Note that we do not explicitly impose any constraints for link or cache capacity in \eqref{Objective_cache}, since they are already incorporated in the cost functions.
We next tackle \eqref{Objective_cache} first in a fixed-routing setting (Section \ref{sec:fixed routing}), and then a general dynamic-routing setting (Section \ref{sec:general condition}).



\section{Special case: fixed-routing}
\label{sec:fixed routing}
The fixed-routing case refers to scenarios where the routing path of a request is fixed or pre-determined. 
Namely, if node $i$ is not a designated server of item $k$, the request packets of $k$ arriving at $i$ can only be forwarded to one pre-defined next-hop neighbor of $i$. 
We denote such next-hop of $i$ for $k$ as $j_{i}(k)$.

\subsection{A DR-submodular + concave reformulation}
In the fixed-routing case, problem \eqref{Objective_cache} reduces to 
\begin{equation}
\begin{aligned}
    \min_{\boldsymbol{y}} \quad & T(\boldsymbol{y}) = \sum\nolimits_{(i,j)\in\mathcal{E}}D_{ij}(F_{ij}) + \sum\nolimits_{i \in \mathcal{V}}B_{i}(Y_{i})
    \\
    \text{subject to} \quad & 0 \leq y_{i} \leq 1, \quad \forall k \in \mathcal{C}, i \not\in \mathcal{S}_k,
    \\ & \phi_{ij}(k) = \begin{cases}
    1 -  y_{i}, \quad \text{if } i \not\in \mathcal{S}_k \text{ and } j =  j_i(k),
    \\ 0, \quad \text{ otherwise.}
    \end{cases}
\end{aligned}
\label{fixroute:Obj}
\end{equation}

Let $p_{vk}$ be the routing path from node $v$ to a designated server $s_k \in \mathcal{S}_k$. 
Path $p_{vk}$ is a node sequence $(p_{vk}^1, p_{vk}^2, \cdots, p_{vk}^{|p_{vk}|})$, where $p_{vk}^1 = v$, $p_{vk}^{|p_{vk}|} = s_k$, and $p_{vk}^{l+1} = j_{p_{vk}^l}(k)$ for $l = 1,\cdots,|p_{vk}|-1$.
We say $(i,j) \in p_{vk}$ for a link $(i,j)$ if $i$ and $j$ are two consecutive nodes in $p_{vk}$.
If $i \in p_{vk}$, let $l_{p_{vk}}(i)$ denote the position of $i$ on path $p_{vk}$, i.e., $p_{vk}^{l_{p_{vk}}(i)} = i$.
We assume every path $p_{vk}$ is well-routed, i.e., no routing loop is formed, and no intermediate node is a designated server of $k$.
Therefore, in terms of item $k$, the rate of request packets that are generated by node $v$ and arrive at node $i$ is given by $ r_{v}(k) \prod_{l^\prime = 1}^{l_{p_{vk}}(i) -1}\left(1 - y_{ p_{vk}^{l^\prime}}(k)\right)$ if $i \in p_{vk}$, and $0$ if $i \not\in p_{vk}$.
Thus,
\begin{align*}
    t_{i}(k) = \sum\nolimits_{v : i \in p_{vk}} r_{v}(k) \prod\nolimits_{l^\prime = 1}^{l_{p_{vk}}(i)-1}\left(1 - y_{ p_{vk}^{l^\prime}}(k)\right).
\end{align*}
Then the link flow rates are given by
\begin{equation}
 \begin{aligned}
     f_{ji}(k) 
     = \sum\nolimits_{v : (i,j) \in p_{vk}} r_{v}(k) \prod\nolimits_{l^\prime = 1}^{l_{p_{vk}}(i)}\left(1 - y_{ p_{vk}^{l^\prime}}(k)\right). \label{fixroute:Fij}
 \end{aligned}
\end{equation}
 

We denote by $T(\boldsymbol{0})$ the cost when $\boldsymbol{y} = \boldsymbol{0}$, i.e., the total routing costs when no cache is deployed, and we assume $T(\boldsymbol{0})$ is finite. 
Then problem \eqref{fixroute:Obj} is equivalent to maximizing a caching gain:
\begin{equation}
\begin{aligned}
   \max_{\boldsymbol{y}} \quad  &G(\boldsymbol{y}) =  A(\boldsymbol{y}) - B(\boldsymbol{y})
\\ \text{subject to} \quad & 0 \leq y_{i}(k) \leq 1, \quad \forall k \in \mathcal{C}, i \not\in \mathcal{S}_k
\end{aligned}
\label{fixroute:max_G}
\end{equation}
where $A(\boldsymbol{y})$ and $B(\boldsymbol{y})$ are given by
\begin{equation*}
\begin{aligned}
A(\boldsymbol{y}) = T(\boldsymbol{0}) - \sum\nolimits_{(i,j) \in \mathcal{E}}D_{ij}(F_{ij}), \quad B(\boldsymbol{y}) = \sum\nolimits_{i \in \mathcal{V}}B_i(Y_i).
\end{aligned}
\end{equation*}

Formulation \eqref{fixroute:max_G} can be viewed as an extension of \cite{mahdian2020kelly} with elastic cache sizes.
In fact, function $G(\boldsymbol{y})$ falls into the category of ``DR-submodular $+$ concave'', the proof is provided in Appendix \ref{proof_lemma_submodular}.

\begin{lem}
 Problem \eqref{fixroute:max_G} is a ``DR-submodular + concave'' maximization problem. Specifically, $A(\boldsymbol{y})$ is non-negative monotonic DR-submodular
 \footnote{DR-submodular function is a continuous generalization of submodular functions with diminishing return. We refer the readers to \cite{bian2017guaranteed} and Appendix \ref{appendix_non_DRsubmodular} for more information.}
 in $\boldsymbol{y}$, and $B(\boldsymbol{y})$ is convex in $\boldsymbol{y}$.
\label{lemma:fixroute:submodular}
\end{lem}

\subsection{Algorithm with (1/2, 1) guarantee}

DR-submodular $+$ concave maximization problem is first systematically studied recently by Mitra et al. \cite{mitra2021submodular}.
Problem \eqref{fixroute:max_G} falls into one of the categories in \cite{mitra2021submodular}, where a Gradient-combining Frank-Wolfe algorithm (Algorithm \ref{alg_GCFW}) guarantees a $(\frac{1}{2}, 1)$ approximation.

\begin{theo}[Theorem 3.10 \cite{mitra2021submodular}]
\label{theorem:fixroute:GCFW_guarantee}
We assume $G$ is L-smooth, i.e., $\nabla G$ is Lipschitz continuous.
For $N > 1$, let $\boldsymbol{y}^*$ be an optimal solution to problem \eqref{fixroute:max_G}, then it holds that
\begin{equation*}
    G(\boldsymbol{y}^{\text{out}}) \geq \frac{1 - \varepsilon}{2} A(\boldsymbol{y}^*) - B(\boldsymbol{y}^*) - \varepsilon \cdot O\left(L |\mathcal{V}||\mathcal{C}|\right).
\end{equation*}
where $L$ is the Lipschitz constant
of $\nabla G$.
\end{theo}

\begin{algorithm}[t]
\SetKwRepeat{DoFor}{do}{for}
\SetKwRepeat{DoDuring}{do}{during}
\SetKwRepeat{DoAt}{do}{at}
\SetKwRepeat{DoWhen}{do}{when}
\SetKwInput{KwInput}{Input}
\KwInput{Integer $N > 1$}
\KwResult{Cache strategy $\boldsymbol{y}^{\text{out}}$ for fixed-routing case}
Start with $n = 0$, let $\varepsilon = N^{-\frac{1}{3}}$.\\
Set $\boldsymbol{y}^{(0)}$ to be $y_i(k) = 0$ for all $i$,$k$.\\
\DoFor{ $n = 0,1,\cdots, N-1$}
{
Let $\boldsymbol{s}^{(n)} = \arg\max_{\boldsymbol{0} \leq \boldsymbol{y} \leq \boldsymbol{1}}\, \left\langle \boldsymbol{y}, \nabla A\left(\boldsymbol{y}^{(n)}\right) -2 \nabla B(\boldsymbol{y}^{(n)}) \right\rangle$. \label{line_Linear_Programming} \\
Let $\boldsymbol{y}^{(n+1)} = (1 - \varepsilon^2) \boldsymbol{y}^{(n)} + \varepsilon^2 \boldsymbol{s}^{(n)}$.\\
}
Find the best among $\left\{ \boldsymbol{y}^{(0)},\cdots,\boldsymbol{y}^{(N)}\right\}$, let $\boldsymbol{y}^{\text{out}} = \arg\max_{\boldsymbol{y} \in \left\{ \boldsymbol{y}^{(0)},\cdots,\boldsymbol{y}^{(N)}\right\}}\, G(\boldsymbol{y})$ .\\
\caption{Gradient-combining Frank-Wolfe (GCFW)}
\label{alg_GCFW}
\end{algorithm}

By \eqref{fixroute:Fij}, the gradient $\nabla B(\boldsymbol{y})$ in Algorithm \ref{alg_GCFW} can be calculated as $\frac{\partial B(\boldsymbol{y})}{\partial y_{z}(k)} = B^\prime_{z}(Y_{z})$, and $\nabla A(\boldsymbol{y})$ is given by 
\begin{equation*}
\begin{gathered}
\frac{\partial A(\boldsymbol{y})}{\partial y_{z}(k)} 
=  t_z(k)\sum\nolimits_{(i,j) \in p_{zk}} D^{\prime}_{ji}(F_{ji})  \prod\nolimits_{l^\prime = 2}^{l_{p_{zk}}(i)}\left(1 - y_{p_{zk}^{l^\prime}}(k)\right).
\end{gathered}
\end{equation*}
The linear programming in Algorithm \ref{alg_GCFW} can be implemented simply by selecting node $z$ and item $k$ with $\frac{\partial A(\boldsymbol{y})}{\partial y_z(k)} - 2\frac{ \partial B(\boldsymbol{y})}{\partial y_z(k)} > 0$
and let elements in $\boldsymbol{s}^{(n)}$ for these $z$ and $k$ be $1$, while keeping others $0$.

\section{General case: dynamic-routing}
\label{sec:general condition}

The analysis in Section \ref{sec:fixed routing} is not applicable to the dynamic-routing case, because the DR-submodularity no longer holds with flexible routing\footnote{We provide in Appendix \ref{appendix_non_DRsubmodular} a detailed explanation to such loss of DR-submodularity.}.
In this section, we tackle the general case with a node-based perspective first used in \cite{gallager1977minimum} and followed by \cite{mahdian2018mindelay,wiopt22}.
We first present a KKT necessary optimality condition for \eqref{Objective_cache}, then give a modification to the KKT condition.
We show that the modified condition yields a bounded gap from the global optimum, then provide further discussion and corollaries.

\subsection{KKT necessary condition}
\label{subsection:KKT necessary condition}
Following \cite{gallager1977minimum}, we start by giving closed-form partial derivatives of $T(\boldsymbol{\phi},\boldsymbol{y})$.
For caching strategy $\boldsymbol{y}$, it holds that $\frac{\partial T}{\partial y_{i}(k)} = B^\prime_{i}(Y_{i})$.

For routing strategy $\boldsymbol{\phi}$, 
the marginal cost due to increase of $\phi_{ij}(k)$ equals a sum of two parts, (1) the marginal cost due to increase of $F_{ji}$ since more responses are sent from $j$ to $i$, and (2) the marginal cost due to increase of $r_j(k)$ since node $j$ needs to handle more request packets.
Formally,
\begin{equation}
    \frac{\partial T}{\partial \phi_{ij}(k)} = t_i(k)\left(D^\prime_{ji}(F_{ji}) + \frac{\partial T}{\partial r_j(k)}\right),
    \label{pT_pphi_cache}
\end{equation}
where the term $\partial T/\partial r_i(k)$ is the marginal cost for $i$ to handle unit rate increment of request packets for $k$, and equals a weighted sum of marginal costs on out-going links and neighbors. Namely,
\begin{equation}
    \frac{\partial T}{\partial r_i(k)} = \sum\nolimits_{j \in \mathcal{N}(i)} \phi_{ij}(k) 
    \left(D^\prime_{ji}(F_{ji}) + \frac{\partial T}{\partial r_j(k)}\right).
    \label{pT_pr_cache}
\end{equation}

By constraint \eqref{FlowConservation_cache}, the value of $\partial T/\partial r_i(k)$ is implicitly affected by $y_i(k)$, e.g., it holds that $\partial T/\partial r_i(k) = 0$ if $i \in \mathcal{S}_k$ or $y_i(k) = 1$.\footnote{
If no routing loops are formed, $\partial T/\partial r_i(k)$ can be computed recursively by \eqref{pT_pr_cache}, staring from nodes $i \in \mathcal{S}_k$ or with $y_i(k) = 1$.
}


\begin{theo}
Let $(\boldsymbol{\phi},\boldsymbol{y})$ be an optimal solution to problem \eqref{Objective_cache}, then for any $i \in \mathcal{V}$ and $k \in \mathcal{C}$,
\begin{equation}
    \begin{aligned}
    B^\prime_{i}(Y_{i})
        &\begin{cases}
            = \lambda_{ik}, \quad \text{if } y_{i}(k) > 0,
            \\ \geq \lambda_{ik}, \quad \text{if } y_{i}(k) = 0,
        \end{cases}
        \\
        t_i(k)\left(D^\prime_{ji}(F_{ji}) + \frac{\partial T}{\partial r_j(k)}\right)
        &\begin{cases}
            = \lambda_{ik}, \quad \text{if } \phi_{ij}(k) > 0,
            \\ \geq \lambda_{ik}, \quad \text{if } \phi_{ij}(k) = 0,
        \end{cases} \quad \forall j \in \mathcal{N}(i).
    \end{aligned}
\label{condition_necessary_cache}
\end{equation}
where $\lambda_{ik}$ is given by
\begin{equation}
\begin{aligned}
    \lambda_{ik} 
    = \min\left\{ B^\prime_{i}(Y_{i}) , \min_{j \in \mathcal{N}(i)} t_i(k)\left(D^\prime_{ji}(F_{ji}) + \frac{\partial T}{\partial r_{j}(k)}\right)\right\}.
\end{aligned}
\label{necessary_condition_lambda}
\end{equation}
\label{thm_necessary_cache}
\end{theo}
Theorem \ref{thm_necessary_cache} gives a KKT necessary condition for problem \eqref{Objective_cache}.
The proof is provided in Appendix \ref{proof_thm_necessary}.
Note that Condition \eqref{condition_necessary_cache} is not sufficient for global optimality even for the pure-routing problem, i.e., with $\boldsymbol{y}$ fixed to $\boldsymbol{0}$. A counterexample is provided in \cite{gallager1977minimum}.
Such non-sufficiency is caused by the degenerate case (i.e., saddle points) where $t_i(k) = 0$, in which $\lambda_{ik}$ is always $0$ and \eqref{condition_necessary_cache} always holds, regardless of routing strategies $[\phi_{ij}(k)]_{j \in \mathcal{V}}$.

\subsection{Modified condition}
\label{subsec:revised condition}
We next propose a modification to \eqref{condition_necessary_cache} that removes the degenerate case at $t_i(k) = 0$.
Note that when $t_i(k) = 0$, it is optimal to set $y_i(k) = 0$ since no request packets for item $k$ ever arrive at node $i$.
Note also that $t_i(k)$ appears repeatedly in \eqref{necessary_condition_lambda} for all $j \in \mathcal{N}(i)$.
We therefore divide all terms in \eqref{necessary_condition_lambda} by $t_i(k)$ and arrive at condition \eqref{condition_BoundedGap_cache}.
A bounded gap on the total cost is promised by \eqref{condition_BoundedGap_cache}.


\begin{theo}
Let $(\boldsymbol{\phi},\boldsymbol{y})$ be feasible to problem \eqref{Objective_cache}, such that for all $i \in \mathcal{V}$ and $k \in \mathcal{C}$,
\allowdisplaybreaks
\begin{equation}
\begin{aligned}
    B^\prime_{i}(Y_{i})
        &\begin{cases}
            = t_i(k)\delta_{i}(k), \quad \text{if } y_{i}(k) > 0,
            \\ \geq t_i(k)\delta_{i}(k), \quad \text{if } y_{i}(k) = 0,
        \end{cases} 
    \\D^\prime_{ji}(F_{ji}) + \frac{\partial T}{\partial r_j(k)}
        &\begin{cases}
            = \delta_{i}(k), \quad \text{if } \phi_{ij}(k) > 0,
            \\ \geq \delta_{i}(k), \quad \text{if } \phi_{ij}(k) = 0,
        \end{cases} \quad \forall j \in \mathcal{N}(i),
\end{aligned}
\label{condition_BoundedGap_cache}
\end{equation}
where $\delta_{i}(k)$ is given by \footnote{In the calculation of $\delta_{i}(k)$, we assume $B^\prime_{i}(Y_{i})/t_i(k) = \infty$ if $t_{i}(k) = 0$.}
\begin{equation}
    \delta_{i}(k) = \min\left\{ \frac{B^\prime_{i}(Y_{i})}{t_i(k)}, \min_{j \in \mathcal{N}(i)} \left(D^\prime_{j i}(F_{j i}) + \frac{\partial T}{\partial r_{j}(k)}\right) \right\}.
\label{delta_BoundedGap_cache}
\end{equation}

Let $(\boldsymbol{\phi}^\dagger,\boldsymbol{y}^\dagger)$ be any feasible solution to \eqref{Objective_cache}. 
Then, it holds that
\begin{equation}
\begin{gathered}
    T(\boldsymbol{\phi}^\dagger,\boldsymbol{y}^\dagger) - T(\boldsymbol{\phi},\boldsymbol{y}) \geq 
    \\\sum\nolimits_{i \in \mathcal{V}}\sum\nolimits_{k \in \mathcal{C}}\delta_{i}(k)\left(y_{i}(k) - y_i^\dagger(k)\right)\left(t_i^\dagger(k)-t_i(k)\right).
\end{gathered}
\label{gap_BoundedGap_cache}
\end{equation}
\label{thm_BoundedGap_cache}
\end{theo}
The proof is provided in Appendix \ref{proof_bounded_gap}. { We remark that function $T(\boldsymbol{\phi},\boldsymbol{y})$ is a summation of a convex function and a geodesic convex function. To the best of our knowledge, we are the first to formulate and tackle such problem with a provable bound. See Appendix \ref{proof_bounded_gap} for more detail.}
Note that \eqref{condition_BoundedGap_cache} implies $y_{i}(k) = 0$ if $t_i(k) = 0$, since the increasing and convex assumption of $B_{i}(\cdot)$ requires $B^\prime_{i}(Y_{i}) > 0$ if $Y_{i} > 0$.
Condition \eqref{condition_BoundedGap_cache} is a more restrictive version of the necessary condition \eqref{condition_necessary_cache}.
Any feasible $(\boldsymbol{\phi},\boldsymbol{y})$ satisfying \eqref{condition_BoundedGap_cache} must also satisfy \eqref{condition_necessary_cache}.
Unlike \cite{gallager1977minimum,wiopt22}, condition \eqref{condition_BoundedGap_cache} is still not sufficient for global optimality. 
Nevertheless, it is practically efficient to minimize the total cost in a distributed manner according to \eqref{condition_BoundedGap_cache}. We next provide further discussion upon condition \eqref{condition_BoundedGap_cache}.

\subsection{Intuitive interpretation}
To provide an intuitive interpretation of the modified condition, 
let $\delta_{ij}(k)$ denote the marginal cost due to increase of flow rate $f_{ji}(k)$, that is, the marginal cost if node $i$ forwards additional requests of unit rate to node $j$.
Then similar to \eqref{pT_pphi_cache} and \eqref{pT_pr_cache}, $\delta_{ij}(k)$ is given by
\begin{equation}
    \delta_{ij}(k) = \frac{\partial T}{\partial f_{ij}(k)}
    = D^\prime_{ji}(F_{ji}) + \frac{\partial T}{\partial r_j(k)}. 
\label{delta_ij_intuitive}
\end{equation}

On the other hand, we define a virtual \emph{cached flow} as $f_{i0}(k) = t_i(k) y_i(k)$, i.e., the rate of request packets for item $k$ that terminate at node $i$ due to $i$'s caching strategy.
Let $\delta_{i0}(k)$ denote the marginal cost due to increase of $f_{i0}(k)$, namely,
\begin{equation}
    \delta_{i0}(k) = \frac{\partial T}{\partial f_{i0}(k)}
    = \frac{\partial T}{t_i(k)\partial y_i(k)} = \frac{B^\prime_i(Y_i)}{t_i(k)}.
\label{delta_i0_intuitive}
\end{equation}
By \eqref{FlowConservation_cache}, $\delta_{i0}(k)$ gives the marginal cache deployment cost if $i$ wishes to increase $y_i(k)$ so that the total request packets forwarded to its neighbors is reduced by unit rate.
Therefore, by \eqref{delta_BoundedGap_cache}, we have
\begin{equation}
    \delta_i(k) = \min_{j \in \left\{0\right\} \cup \mathcal{V}} \delta_{ij}(k).
\label{delta_ik_intuitive}
\end{equation}
That is, $\delta_i(k)$ gives the minimum marginal cost for node $i$ to handle request packets for item $k$.
Theorem \ref{thm_BoundedGap_cache} then suggests that each node handles incremental arrival requests in the way that achieves its minimum marginal cost -- either by forwarding to neighbors, or by expanding its own cache.
In other words, it is ``worthwhile'' to deploy cache for $k$ at $i$ if $\delta_{i0}(k) < \min_{j \in \mathcal{N}} \delta_{ij}(k)$, and ``not worthwhile'' otherwise.


\subsection{Corollaries}
We provide a few corollaries to further investigate condition \eqref{condition_BoundedGap_cache}.
\begin{cor}
For any optimal solution $(\boldsymbol{\phi}^*,\boldsymbol{y}^*)$ to \eqref{Objective_cache}, there must exist a corresponding $(\boldsymbol{\phi},\boldsymbol{y})$ satisfying condition \eqref{condition_BoundedGap_cache}, such that $\boldsymbol{y} = \boldsymbol{y}^*$ and $\phi_{ij}(k) = \phi^*_{ij}(k)$ for all $i,k$ with $t_i^*(k) > 0$.
\label{cor_cache_existance}
\end{cor}
The proof is provided in Appendix \ref{proof_cor_existance}.
Corollary \ref{cor_cache_existance} implies that condition \eqref{condition_BoundedGap_cache} must have non-empty intersection with the global optima of \eqref{Objective_cache}, even though it is neither a necessary condition nor a sufficient condition.

\begin{figure}[htbp]
\centerline{\includegraphics[width=0.9\linewidth]{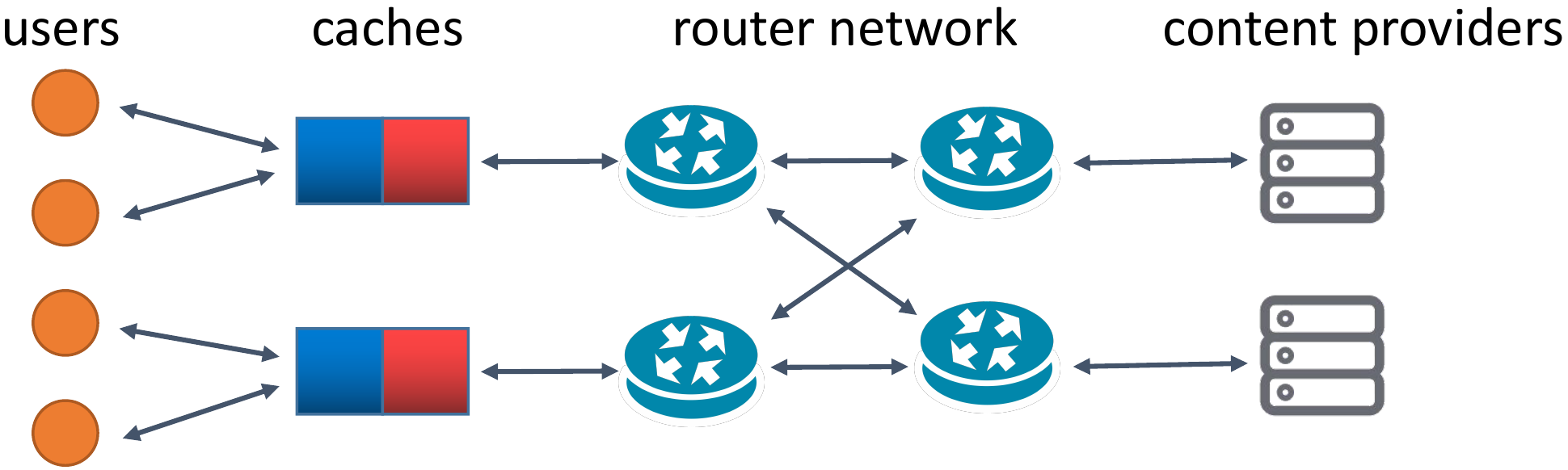}}
\caption{A special scenario that \eqref{condition_BoundedGap_cache} yields a global optimal solution. Single-layered caches are equipped near users. Requests are routed to servers if not fulfilled at the caches.}
\label{fig_opt_case}
\end{figure}

\begin{cor}
Let $(\boldsymbol{\phi},\boldsymbol{y})$ be a feasible solution to \eqref{Objective_cache} and satisfy \eqref{condition_BoundedGap_cache}.
Let $(\boldsymbol{\phi}^\dagger,\boldsymbol{y}^\dagger)$ be a feasible solution to \eqref{Objective_cache}, such that for all $i \in \mathcal{V}$ and $k \in \mathcal{C}$, either $y_{i}^\dagger(k) = y_{i}(k)$ or $t_{i}^\dagger(k) = t_{i}(k)$.
Then it holds that $T(\boldsymbol{\phi},\boldsymbol{y}) \leq T(\boldsymbol{\phi}^\dagger,\boldsymbol{y}^\dagger)$.
\label{cor_cache_1}
\end{cor}
Corollary \ref{cor_cache_1} is obvious from Theorem \ref{thm_BoundedGap_cache}.
It implies that \eqref{condition_BoundedGap_cache} is sufficient for optimal $\boldsymbol{\phi}$ when $\boldsymbol{y}$ is fixed, and for optimal $\boldsymbol{y}$ when $t_i(k)$ are unchanged.
An example is shown in Figure \ref{fig_opt_case}, where the caches always receive the same amount of request packets (i.e., unchanged $t_i(k)$), and the routers can not cache at all (i.e., unchanged $y_i(k)$).

\begin{cor}
Let $(\boldsymbol{\phi},\boldsymbol{y})$ be a feasible solution to \eqref{Objective_cache} and satisfy \eqref{condition_BoundedGap_cache}.
Let $(\boldsymbol{\phi}^\dagger,\boldsymbol{y}^\dagger)$ be a feasible solution to \eqref{Objective_cache}, 
such that either $\boldsymbol{\phi}^\dagger \geq \boldsymbol{\phi}$ or $\boldsymbol{\phi}^\dagger \leq \boldsymbol{\phi}$.\footnote{ For two vectors $\boldsymbol{v}_1$ and $\boldsymbol{v}_2$ of the same dimension, we denote by $\boldsymbol{v}_1 \geq \boldsymbol{v}_2$ if every element of $\boldsymbol{v}_1$ is no less than the corresponding element in $\boldsymbol{v}_2$. Similarly as $\boldsymbol{v}_1 \leq \boldsymbol{v}_2$.}
Then it holds that $T(\boldsymbol{\phi},\boldsymbol{y}) \leq T(\boldsymbol{\phi}^\dagger,\boldsymbol{y}^\dagger)$.
\label{cor_cache_2}
\end{cor}
The proof is provided in Appendix \ref{proof_cor_f_increase}.
For $i,k$ such that $y_{i}(k) \neq 1$, let $\rho_{ij}(k) = \phi_{ij}(k) / \left(1 - y_i(k)\right)$ be the \emph{conditional routing variable}, i.e., the probability of a request packet being forwarded to $j$ given the requested item is not cached at $i$.
In practical networks, the routing and caching mechanisms are usually implemented separately, and the routing is only based on $\rho_{ij}(k)$ instead of $\phi_{ij}(k)$.
Corollary \ref{cor_cache_2} contains a special case where $\rho_{ij}^\dagger(k) = \rho_{ij}(k)$ for all $i,j,k$, but $y_{i}(k)^\dagger \geq y_{i}(k)$ for all $i,k$ (or $y_{i}(k)^\dagger \leq y_{i}(k)$ for all $i,k$).
This special case implies that the total cost cannot be lowered by only caching more items (i.e., only increasing $y_{i}(k)$ for some $i$ and $k$), or only removing items from caches (i.e., only decreasing $y_{i}(k)$ for some $i$ and $k$), while keep the conditional routing variables unchanged.  

\section{Online algorithm}
\label{Sec:GeneralAlgorithm}
We next present a distributed online algorithm for the general case that converges to a loop-free version of the modified condition \eqref{condition_BoundedGap_cache}.
The algorithm does not require prior knowledge of exogenous request rates $r_i(k)$ and designated servers $\mathcal{S}_k$, and is adaptive to moderate changes in $r_i(k)$ and cost functions $D_{ij}(\cdot)$, $B_i(\cdot)$.

\subsection{Algorithm overview}

\begin{figure}[htbp]
\centerline{\includegraphics[width=0.8\linewidth]{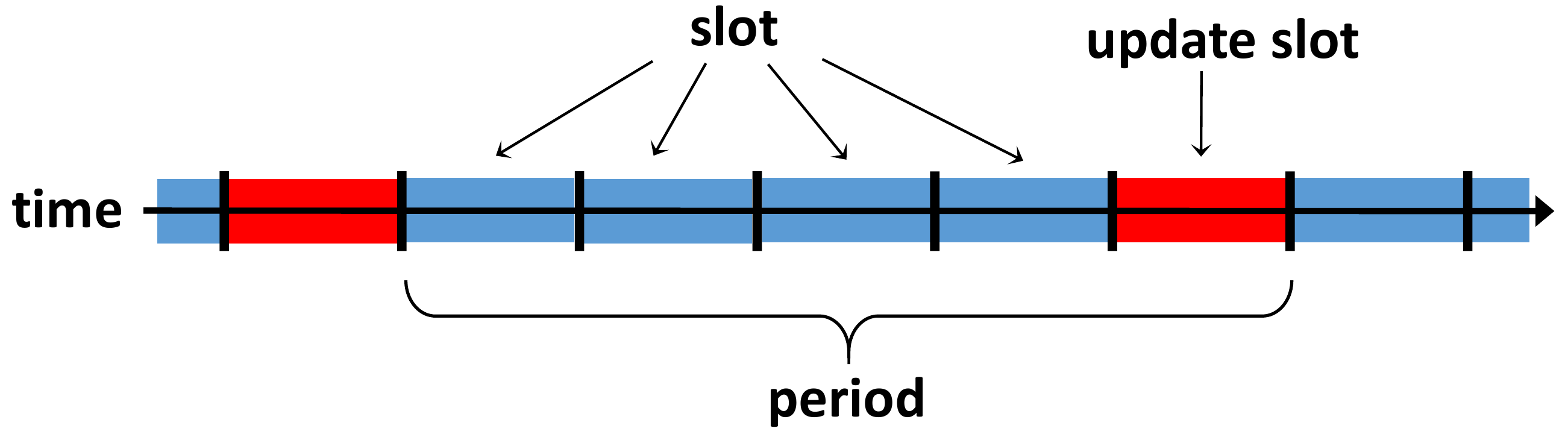}}
\caption{ An example timeline with $L = 5$ slots per period.} 
\label{fig_timeline}
\end{figure}

We partition time into \emph{periods} of duration $LT_{\text{slot}}$. A period consists of $L$ \emph{slots}, each of duration $T_{\text{slot}}$.
In $t$-th period, node $i$ keeps its routing and caching strategies  $(\boldsymbol{\phi}_i^t,\boldsymbol{y}_i^t)$ unchanged.
At the $m$-th slot of $t$-th period ($1 \leq m \leq L$), node $i$ rounds $\boldsymbol{y}_i^t$ into integer caching decisions $\boldsymbol{x}_i^{t,m}$ with $\mathbb{E}[\boldsymbol{x}_i^{t,m}] = \boldsymbol{y}_i^{t}$. \footnote{We suggest refreshing caching decisions multiple times in each period to better estimate theoretical costs and marginals from actual measurements. Nevertheless, the algorithm applies to any $L\geq 1$.}
The last slot in each period is called \emph{update slot}, during which nodes update their routing and caching strategies in a distributed manner.
The time partitioning scheme is illustrated in Figure \ref{fig_timeline}.
We postpone the discussion of randomized rounding techniques to Section \ref{section:rounding}, and now focus on the update of strategies $\boldsymbol{\phi}^t$ and $\boldsymbol{y}^t$.

Our algorithm is a gradient projection variant inspired by \cite{gallager1977minimum}.
Each node updates its strategies in the update slot of $t$-th period,
\begin{equation}
\begin{aligned}
 \boldsymbol{\phi}_i^{t+1} = \boldsymbol{\phi}_i^{t} + \Delta\boldsymbol{\phi}_i^{t},
     \quad \boldsymbol{y}_i^{t+1} = \boldsymbol{y}_i^{t} + \Delta \boldsymbol{y}_i^{t}.
\end{aligned}
\label{variable_update}
\end{equation}
The update vectors $\Delta\boldsymbol{\phi}_i^{t}$ and $\Delta\boldsymbol{y}_i^{t}$ are calculated by
\begin{small}
\begin{equation}
\begin{aligned}
    \Delta\phi_{ij}^{t}(k) &= \begin{cases}
    - \phi_{ij}^{t}(k), & \text{if } j \in \mathcal{B}_i^t(k)
    \\ -\min\left\{ \phi_{ij}^t(k), \alpha e_{ij}^t(k) \right\}, & \text{if } j \in \mathcal{N}(i) \backslash \mathcal{B}_i^t(k) \text{, } e_{ij}^t(k) > 0
    \\ S_i^t(k) / N_i^t(k),
    & \text{if } j \in \mathcal{N}(i) \backslash \mathcal{B}_i^t(k) \text{, } e_{ij}^t(k) = 0
    \end{cases}
    \\ \Delta y_{i}^{t}(k) &= \begin{cases}
    -\min\left\{ y_{i}^t(k), \alpha e_{i0}^t(k) \right\}, &\, \text{if } e_{i0}^t(k) > 0
    \\ S_i^t(k) / N_i^t(k),
    &\,  \text{if } e_{i0}^t(k) = 0
    \end{cases}
\end{aligned}
\label{dphi_and_dy}
\end{equation}
\end{small}
where $\mathcal{B}_i^t(k)$ is the set of \emph{blocked nodes} to suppress routing loops (see Section \ref{sec:blocked set} for a detailed discussion of node blocking mechanism), $\alpha$ is the stepsize
, and\footnote{ $\mathbbm{1}_A$ is the indicator function of event $A$. i.e., $\mathbbm{1}_A = 1$ if $A$ is true, and $0$ if not.}
\begin{small}
\begin{equation}
\begin{aligned}
     &e_{i0}^t(k) = \delta_{i0}^t(k) - \delta_i^t(k),
     \quad e_{ij}^t(k) = \delta_{ij}^t(k) - \delta_i^t(k), \, \forall j \in \mathcal{N}(i) \backslash \mathcal{B}_i^t(k),
     \\ & N_i^t(k) = \bigg|\left\{j \in \mathcal{N}(i) \backslash\mathcal{B}_i^t(k) \big| e_{ij}^t(k) = 0 \right\}\bigg| + \mathbbm{1}_{\delta_{i0}^t(k) > 0},
     \\ & S_i^t(k)= \sum\nolimits_{j \in \mathcal{N}(i) \backslash \mathcal{B}_i^t(k) \,:\, e_{ij}^t(k) > 0}\Delta\phi_{ij}^{t}(k) + \Delta y_{i}^{t}(k)\mathbbm{1}_{\delta_{i0}^t(k) > 0},
\end{aligned}    
\label{algorithm_detail_eNS}
\end{equation}
\end{small}
The intuitive idea is to transfer routing/caching fractions from non-minimum-marginal directions to the minimum-marginal ones.
$\delta_{ij}^t(k)$ and $\delta_{i0}^t(k)$ are calculated as in \eqref{delta_ij_intuitive} and \eqref{delta_i0_intuitive}.
But slightly different from \eqref{delta_ik_intuitive}, due to the existence of $\mathcal{B}_i^t(k)$, $\delta_i^t(k)$ is given by
\begin{equation}
    \delta_{i}^t(k) = \min\left\{ \delta_{i0}^t(k), \min\nolimits_{j \in \mathcal{N}(i) \backslash\mathcal{B}_i^t(k)} \delta_{ij}^t(k) \right\}.
\label{delta_ik_t}
\end{equation}

In each update slot, to calculate $\delta_{ij}^t(k)$ and $\delta_{i0}^t(k)$, the value $\partial T/\partial r_i(k)$ is updated throughout the network with a control message broadcasting mechanism (see, e.g., \cite{wiopt22}). Specifically, node $i$ receives $\partial T/ \partial r_j(k)$ from all downstream neighbors (i.e., the nodes $j \in \mathcal{N}(i)$ with $\phi_{ij}(k) > 0$), calculates\footnote{Node $i$ needs to know the analytical forms of $B_i(\cdot)$ and $D_{ij}(\cdot)$, or be able to estimate $B_i^\prime(Y_i)$ and $D_{ij}^\prime(F_{ij})$ from corresponding cache size $Y_i$ and flow rate $F_{ij}$. 
The flow rate $F_{ij}$ is measured by the average rate during the first $(L-1)$ slots of $t$-th period.} 
its $\partial T/\partial r_i(k)$ according to \eqref{pT_pr_cache}, and broadcasts $\partial T/\partial r_i(k)$ to all upstream neighbors.
Such broadcast starts at the designated servers or nodes with $y_i(k) = 1$, where $\partial T / \partial r_i(k) = 0$.
The proposed algorithm is summarized in Algorithm \ref{alg_GP}.
Next we discuss the set $\mathcal{B}_i^t(k)$.

\begin{algorithm}[t]
\SetKwRepeat{DoFor}{do}{for}
\SetKwRepeat{DoDuring}{do}{during}
\SetKwRepeat{DoAt}{do}{at}
\SetKwRepeat{DoWhen}{do}{when}
\SetKwInput{KwInput}{Input}
\KwInput{Initial loop-free $(\boldsymbol{\phi}^0,\boldsymbol{y}^0)$ with $T^0 < \infty$, stepsize $\alpha$}
Start with $t=0$.\\
\DoAt{ beginning of $m$-th slot of $t$-th period }
{
Each node $i$ round $\boldsymbol{y}_i^t$ into $\boldsymbol{x}_i^{t,m}$ with Distributed Randomized Rounding (\texttt{DRR}).
}

\DoDuring{ update slot of $t$-th period}
{
Each node updates $\partial T/\partial r_i(k)$ for all $k$ via a a message broadcasting mechanism.\\
Each node calculates \eqref{algorithm_detail_eNS}.\\
Each node updates strategies $(\boldsymbol{\phi}_i^t,\boldsymbol{y}_i^t)$ by \eqref{variable_update} and \eqref{dphi_and_dy}.\\
}
\caption{Gradient Projection (GP)}
\label{alg_GP}
\end{algorithm}

\subsection{Loops and blocked nodes}
\label{sec:blocked set}
A routing loop refers to node sequence $(l_1,l_2,\cdots,l_{|l|})$, such that $l_1 = l_{|l|}$ and for some $k \in \mathcal{C}$,  $\phi_{l_p l_{p+1}}(k) > 0$ for all $p = 1,\cdots,|l|-1$.
Such a loop implies that a strictly positive portion of requests for item $k$ forwarded from node $l_1$ is sent back to $l_1$ itself.
The existence of loops should be forbidden, as it gives rise to redundant flow circulation and wastes network resources.
Before discussing loop-preventing mechanisms, we remark that the relaxed formulation \eqref{Objective_cache} may yield loops in its optimal solution due to the continuous relaxation from $\boldsymbol{x}$ to $\boldsymbol{y}$. An example is provided in Appendix \ref{routing_loops}. 

Nevertheless, for practical purposes, our algorithm still prevents the formation of loops by employing a method called ``blocked node set'', assuming a loop-free initial state $\boldsymbol{\phi}^0$ is given.
Specifically, during $(t+1)$-th period, node $i$ should not forward any request of item $k$ to nodes in the blocked node set $\mathcal{B}_i^t(k) \subseteq \mathcal{V}$.
The construction of sets $\mathcal{B}_i^t(k)$ falls into two catalogs, i.e., the \emph{static} sets and the \emph{dynamic} sets.
We next introduce both and describe how to implement them in our algorithm.

\noindent\textbf{Static sets.}
The blocked node sets can be pre-determined and kept unchanged throughout the algorithm, i.e.,  $\mathcal{B}_i^t(k) = \mathcal{B}_i(k)$ for all $t \geq 0$.
A directed acyclic subgraph of $\mathcal{G}$ is constructed for every item at the beginning of the algorithm, in which every node has at least one path to a designated server.
We denote the subgraph w.r.t. $k \in \mathcal{C}$ as $\mathcal{G}_{(k)} = (\mathcal{V}, \mathcal{E}_{(k)})$ with $\mathcal{E}_{(k)} \subseteq \mathcal{E}$. 
Then the blocked node sets are constructed as $\mathcal{B}_i(k) = \left\{j \in \mathcal{N}(i) \big| (i,j) \not\in \mathcal{E}_{(k)} \right\}$.

The idea of fixed blocked node set is commonly adopted, e.g., 
in the FIB construction of ICN. 
The subgraphs $\mathcal{G}_{(k)}$ can be calculated efficiently at the network initialization \cite{ioannidis2017jointly}, either in a centralized way (e.g., Bellman-Ford algorithm), or in a distributed manner (e.g., distance-vector protocol).

\noindent\textbf{Dynamic sets.}
The sets $\mathcal{B}_i^t(k)$ can also be dynamically calculated as the algorithm proceeds.
Compared with the fixed case, dynamically determined sets may give nodes more routing options and, therefore, a potentially better performance in terms of total cost.
It requires a more elaborate node blocking mechanism, preferably distributed and efficient.
A classic dynamic node blocking mechanism is invented in \cite{gallager1977minimum} for a multi-commodity routing problem, however, not applicable in our case.
We develop a novel dynamic node blocking method, presented in Appendix \ref{dynamic_blocked_set}. 
The basic idea of our method is to generate total orders among nodes via topological sorting dynamically during each period.

\subsection{Asynchronous convergence}
In practical ad-hoc networks, nodes may have non-perfect synchronization. Algorithm \ref{alg_GP} allows nodes to update variables at any time during the update slot.
The asynchronous convergence of Algorithm \ref{alg_GP} is stated in Theorem \ref{thm_convergence}.
To model such asynchrony, we assume at the $t$-th iteration, only one node $v(t) \in \mathcal{V}$ updates its variables.
Namely, $(\boldsymbol{y}^{t+1}_i,\boldsymbol{\phi}^{t+1}_i) = (\boldsymbol{y}^{t}_i,\boldsymbol{\phi}^{t}_i)$ for all $i \neq v(t)$,
and we let $\mathcal{T}_i = \left\{t \geq 1 \big| v(t) = i\right\}$ denote the iterations for node $i$'s updates.

Since the node blocking mechanism in Section \ref{sec:blocked set} is implemented to suppress loops, Algorithm \ref{alg_GP} may not converge to condition \eqref{condition_BoundedGap_cache}.
Nevertheless, Theorem \ref{thm_convergence} states that the convergence limit of Algorithm \ref{alg_GP} still satisfies a version of \eqref{condition_BoundedGap_cache} with $\mathcal{N}(i)$ replaced by
$\mathcal{N}(i) \backslash \mathcal{B}_i(k)$.
The proof is provided in Appendix \ref{proof_thm_convergence}.

\begin{theo}
\label{thm_convergence}
Assume the network starts at $(\boldsymbol{\phi}^0, \boldsymbol{y}^0)$ with $T^0 < \infty$, and the variables $(\boldsymbol{\phi}^t, \boldsymbol{y}^t)$ are updated asynchronously by Algorithm \ref{alg_GP} with a sufficiently small stepsize $\alpha$. 
Then, if static blocked node sets are used,
 and $ \lim_{t \to \infty} \left|\mathcal{T}_i\right| = \infty$ for all $i \in \mathcal{V}$, the sequence $\left\{(\boldsymbol{\phi}^t, \boldsymbol{y}^t)\right\}_{t = 0}^{\infty}$ converges to a limit point $(\boldsymbol{\phi}, \boldsymbol{y})$, and $(\boldsymbol{\phi}, \boldsymbol{y})$ satisfies \eqref{condition_BoundedGap_cache}, with $\mathcal{N}(i)$ being replaced by $\mathcal{N}(i) \backslash \mathcal{B}_i(k)$.
\end{theo}

\subsection{Communication complexity}
During each update slot, node $i$ should calculate $\partial T / \partial r_i(k)$ for all $k$. This is accomplished by the control message broadcasting mechanism.
When the network size is large, control messages carrying derivative information may take non-negligible time and bandwidth to percolate the whole network. 
The variables of all nodes are updated once every update slot of duration $T_{\text{slot}}$, and every broadcast message is sent once in an update slot. 
Thus there are $|\mathcal{E}|$ transmissions of broadcast messages corresponding to an item in one update slot, and totally $|\mathcal{C}||\mathcal{E}|$ transmissions each update slot, with on average $|\mathcal{C}|/T_{\text{slot}}$ per link$\cdot$second and at most $d_{\text{max}}|\mathcal{C}|$ per node, where $d_{\text{max}}$ is the largest node degree.
The broadcast messages have $O(1)$ size, and can be sent in an out-of-band channel.
Let $t_c$ be the maximum transmission time for a broadcast message, and
$\Bar{h}$ be the maximum hop number for a request path. 
The completion time of broadcast mechanism at each update is at most $\Bar{h}t_c$.

\subsection{Distributed randomized rounding}
\label{section:rounding}
The continuous caching strategy $\boldsymbol{y}$ is rounded to caching decision $\boldsymbol{x}$ in each slot.
The rounding can be done with a naive probabilistic scheme, i.e., $x_i^{t,m}(k)$ is a Bernoulli random variable with $p = y_i^t(k)$. 
However, such heuristic rounding method may generate huge and impractical cache sizes.
Various advanced rounding techniques exist (see \cite{mahdian2020kelly}). If all $Y_i$ are integer, the deterministic \emph{pipage rounding} 
and the randomized \emph{swap rounding} 
guarantee the actual routing cost after rounding is no worse than the relaxed result, while keeping $X_i = Y_i$.
However, such techniques are centralized.

A distributed rounding method is proposed by \cite{ioannidis2018adaptive}, each node independently operates without knowledge of closed-form $T(\boldsymbol{\phi}, \boldsymbol{y})$.
We extend this method to non-integer $Y_i$, and refer it as Distributed Randomized Rounding (\texttt{DRR}).
We describe in Appendix \ref{appendix_rounding} the detail of \texttt{DRR}.
With such rounding algorithm, it is guaranteed that the expected flow rates and cache sizes meet the relaxed value, and the actual cache size at each node is bounded near the expected value.

\begin{lem}
\label{lemma:rounding}
If $\boldsymbol{x}^{t,m}$ are rounded from $\boldsymbol{y}^{t}$ by \texttt{DRR}, then
\begin{equation*}
\begin{aligned}
    &\mathbb{E}[x_{i}^t(k)] = y_{i}^t(k), \quad \forall i\in\mathcal{V}, k \in \mathcal{C},
     \\& \left|\sum\nolimits_{k \in \mathcal{C}}x_{i}^t(k) - \sum\nolimits_{k \in \mathcal{C}}y_{i}^t(k)\right| < 1, \quad \forall i \in \mathcal{V},
    \\& \mathbb{E}[F_{ij}\big|_{(\boldsymbol{\phi}^t, \boldsymbol{x}^{t,m})}] = F_{ij}\big|_{(\boldsymbol{\phi}^t,\boldsymbol{y}^t)}, \quad \forall (i,j) \in \mathcal{E}.
    \end{aligned}
\end{equation*}
\label{Lem_rounding}
\end{lem}
The proof of Lemma \ref{Lem_rounding} is omitted.
We remark that since $D_{ij}(\cdot)$ and $B_{i}(\cdot)$ are convex, combining with Jensen's Inequality, it holds that $\mathbb{E}\left[T(\boldsymbol{\phi}^t,\boldsymbol{x}^{t,m})\right] \geq T\left(\boldsymbol{\phi}^t,\mathbb{E}[\boldsymbol{x}^{t,m}]\right) = T(\boldsymbol{\phi}^t,\boldsymbol{y}^t)$. 
Nevertheless, we demonstrate in Section \ref{sec:result_grid-25} that, with proper randomized packet forwarding mechanism, the costs measured in real network will not deviate too much from the theoretical result $T(\boldsymbol{\phi}^t,\boldsymbol{y}^t)$.

\section{Simulation}
\label{sec:simulation}
We simulate the proposed algorithms and other baseline methods in various network scenarios with a packet-level simulator\footnote{Available at https://github.com/JinkunZhang/Elastic-Caching-Networks.git}.
\begin{figure*}[h]
  \centering
  \includegraphics[width=0.8\linewidth]{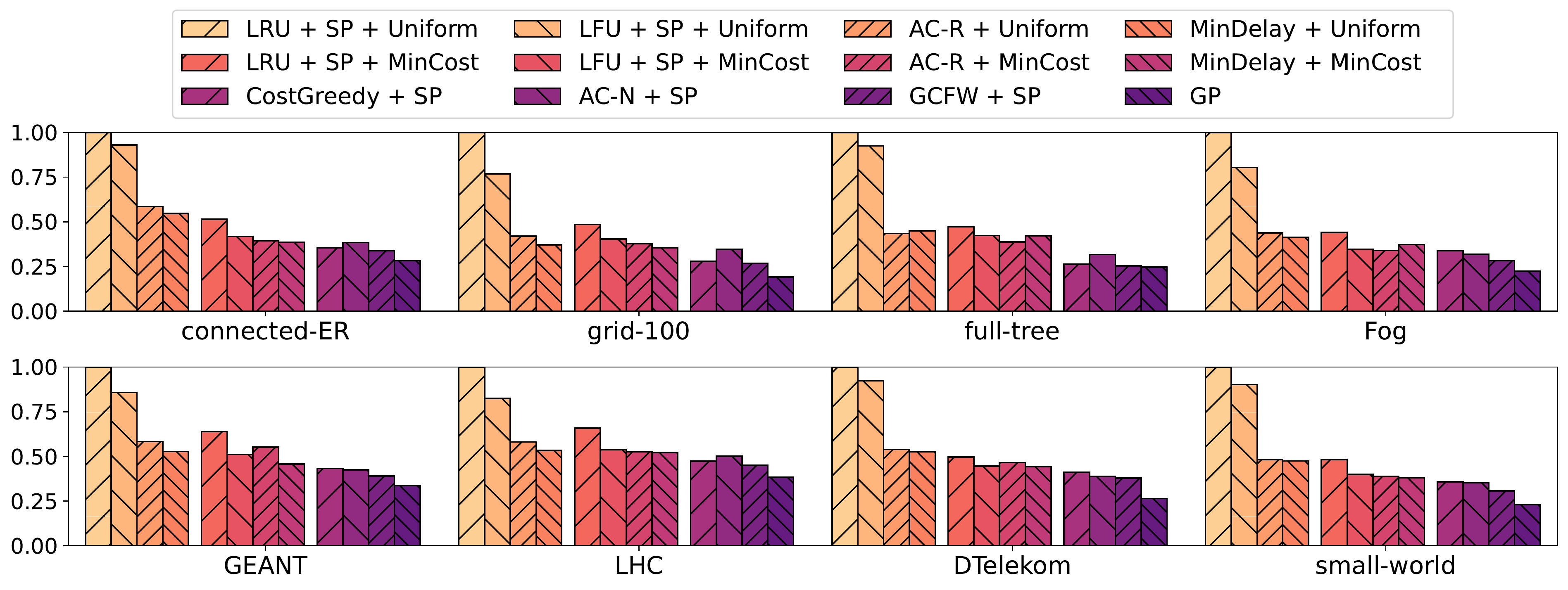}
  \caption{Normalized total cost $T$ of different methods in multiple network scenarios}
 \label{fig_bar1}
\end{figure*}

\subsection{Simulator setting}
\label{sec:simulator setting}
\noindent\textbf{Requests.} 
We denote by $\mathcal{R}$ the set of requests in the network. For each request $(i,k)$, the requester $i$ is uniformly chosen in all $|\mathcal{V}|$ nodes, and the requested item $k$ is chosen in the catalog $\mathcal{C}$ with a Zipf-distribution of parameter $1.0$.
The exogenous request rates $r_{ik}$ for all requests are uniformly random in interval $[1.0, 5.0]$.
For each $(i,k) \in \mathcal{R}$, node $i$ sends request packets for item $k$ in a Poisson process of rate $r_{i}(k)$.
For each item in $k \in \mathcal{C}$, we assume $|\mathcal{S}_k| = 1$ and choose the designated server uniformly randomly in all nodes.

\noindent\textbf{Packet routing.}
The request packets are routed according to variable $\boldsymbol{\phi}$. 
To ensure relatively steady flow rates, we adopt a \emph{token-based} randomized forwarding.
Specifically, every node $i$ keeps a \emph{token pool} for each item $k$, where the tokens represents next-hop nodes $j \in \mathcal{N}(i)$.
A token pool initially has $N_{\text{token}} = 50$ tokens, the number of tokens for node $j$ is in proportion to the corresponding $\phi_{ij}^0(k)$. 
When node $i$ needs to forward a request packet for item $k$, it randomly picks a token from the pool and forwards to the neighbor corresponding to the token. The picked token is then removed from the pool.
When a token pool becomes empty, node $i$ refills it according to the current variable $\boldsymbol{\phi}_i^t$.  

\noindent\textbf{Measurements.}
We monitor the network status at time points with interval $T_{\text{monitor}}$. 
Flows $F_{ij}$ are measured as the average response packets traveled though $(i,j)$ during past $T_{\text{monitor}}$. 
We select a congestion-dependent link cost function $D_{ij}(F_{ij}) = d_{ij}F_{ij} + d_{ij}^2F_{ij}^2 + d_{ij}^3F_{ij}^3$, which is a $3$-order expansion of queueing delay $F_{ij}/ (1/d_{ij} - F_{ij})$. 
For methods only considering linear link costs (e.g., when calculating the shortest path), we use the marginal cost $D_{ij}^\prime(0) = d_{ij}$ for the link weights, representing the unit-flow cost with no congestion.
Cache sizes $Y_i$ are measured as snapshot count of cached items at the monitor time points, and cache deployment cost $B_i(Y_i) = b_i Y_i$, where $b_i$ is the unit cache price at $i$.
Parameters $d_{ij}$ and $b_i$ are uniformly selected from the interval in Table \ref{tab:scenarios}.

\subsection{Simulated scenarios and baselines}
We simulate on multiple synthetic or real-world network scenarios summarized in Table \ref{tab:scenarios}.
\textbf{\texttt{connected-ER}} is a connectivity-guaranteed Erd\H{o}s-R\'enyi graph, where bi-directional links exist for each pair of nodes with probability $p = 0.07$. 
\textbf{\texttt{grid-100}} and \textbf{\texttt{grid-25}} are $2$-dimensional $10\times10$ and $5\times5$ grid networks. 
\textbf{\texttt{full-tree}} is a full binary tree of depth $6$. 
\textbf{\texttt{Fog}} is a full $3$-ary tree of depth $4$, where children of the same parent is concatenated linearly. This topology is dedicated to formulate fog-caching and computing networks \cite{kamran2021deco}.
\textbf{\texttt{GEANT}} is a pan-European data network for the research and education community \cite{rossi2011caching}.
\textbf{\texttt{LHC}} (Large Hadron Collider) is a prominent data-intensive computing network for high energy physics applications.
\textbf{\texttt{DTelekom}} is a sample topology of Deutsche Telekom company \cite{rossi2011caching}.
\textbf{\texttt{small-world}} (Watts-Strogatz small world) is a ring-graph with additional short-range and long-range edges.


\begin{table}
  \caption{Simulated Network Scenarios}
  \begin{tabular}{c | cccccc}
    \toprule
    Topologies & $|\mathcal{V}|$  & $|\mathcal{E}|$ & $|\mathcal{C}|$ & $|\mathcal{R}|$  & $d_{ij}$ & $b_i$  \\
    \midrule
    \texttt{connected-ER} & 50 & 256 & 80 & 200 & [0.05, 0.1] & [5, 10] \\
    \texttt{grid-100} & 100 & 358 & 100  & 400 & [0.05, 0.1] & [20, 40]  \\
\texttt{full-tree} & 63 & 124 & 50  & 150 & [0.05, 0.1] & [20, 30]  \\
\texttt{Fog} & 40 & 130 & 50 & 200 & [0.05, 0.1] & [30, 50]  \\
\texttt{GEANT} & 22 & 66 & 40 & 100 & [0.05, 0.1] & [10, 15]  \\
\texttt{LHC} & 16 & 62 & 30 & 100  & [0.1, 0.15] & [10, 15] \\
\texttt{DTelekom} & 68 & 546 & 100 & 300 & [0.1, 0.2] & [10, 20] \\
\texttt{small-world} & 120 & 720 & 100 & 400 & [0.05, 0.1] & [10, 20]  \\
\texttt{grid-25} & 25 & 80 & 30 &  100 & 0.1 & 10 \\
    \bottomrule
  \end{tabular}
 \label{tab:scenarios}
\end{table}

\begin{table}
  \caption{Implemented methods and functionalities}
  \begin{tabular}{c | ccccc}
    \toprule
    Functionality & \texttt{LRU}/\texttt{LFU} & \texttt{SP} & \texttt{AC-R} & \texttt{MinDelay} & \texttt{Uniform}  \\
    \midrule
    cache deployment & & & & & \checkmark\\
    content placement & \checkmark & & \checkmark & \checkmark & \\
    routing & & \checkmark & \checkmark& \checkmark &\\
    \bottomrule
  \end{tabular}
  \begin{tabular}{c | cccccccccc}
    \toprule
    Functionality & \texttt{MinCost} & \texttt{CostGreedy} & \texttt{AC-N} & \texttt{GCFW} & \texttt{GP} \\
    \midrule
    cache deployment & \checkmark & \checkmark &\checkmark & \checkmark  & \checkmark \\
    content placement & & \checkmark & \checkmark & \checkmark & \checkmark \\
    routing &  & & & & \checkmark\\
    \bottomrule
  \end{tabular}
 \label{tab:methods}
\end{table}

\begin{figure*}[htbp]
\centering
\begin{minipage}[t]{0.38\textwidth}
\includegraphics[width=1\textwidth]{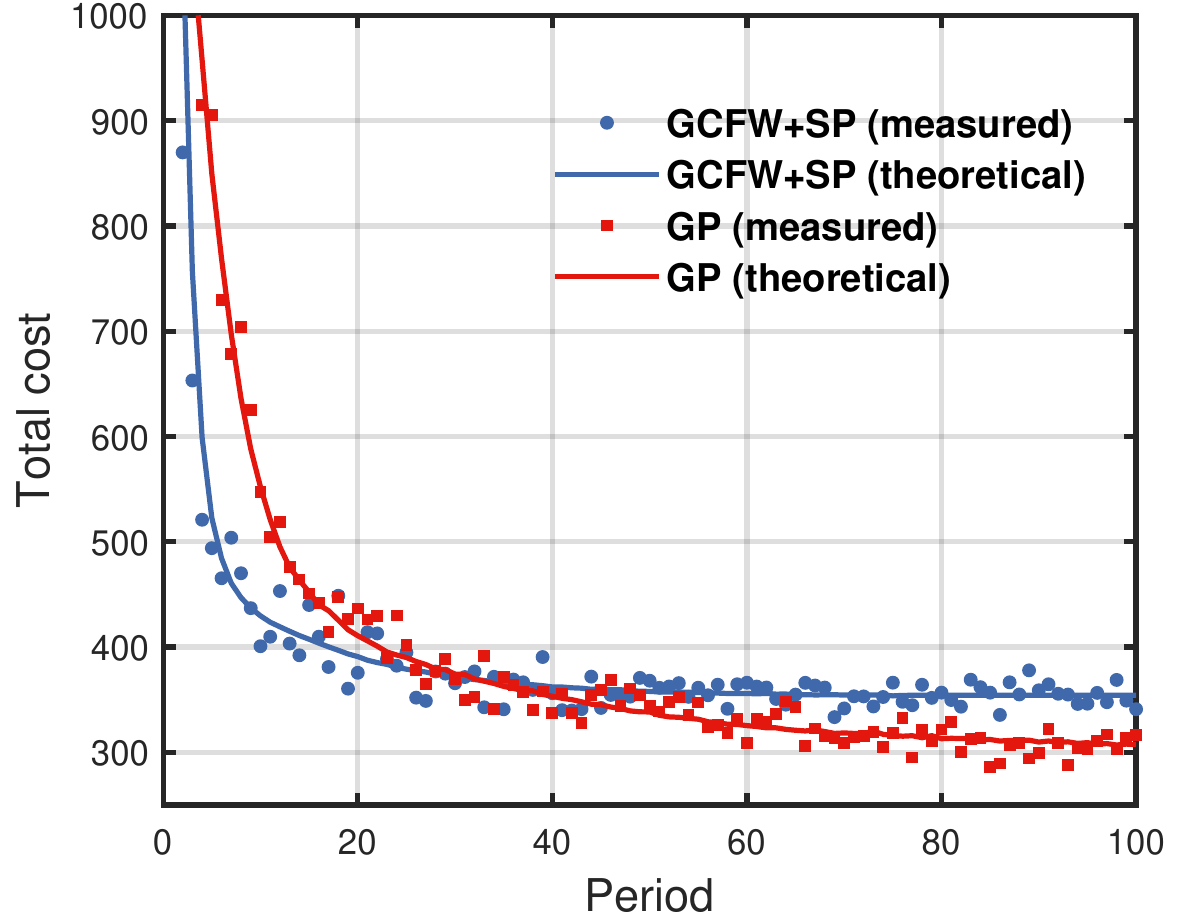}
\caption{Convergence of measured and theoretical total costs by \texttt{GCFW}+\texttt{SP} and \texttt{GP} in \texttt{grid-25}}
\label{fig_convergence}
\end{minipage}
\hspace{2mm}
\begin{minipage}[t]{0.38\textwidth}
\centering
\includegraphics[width=1\textwidth]{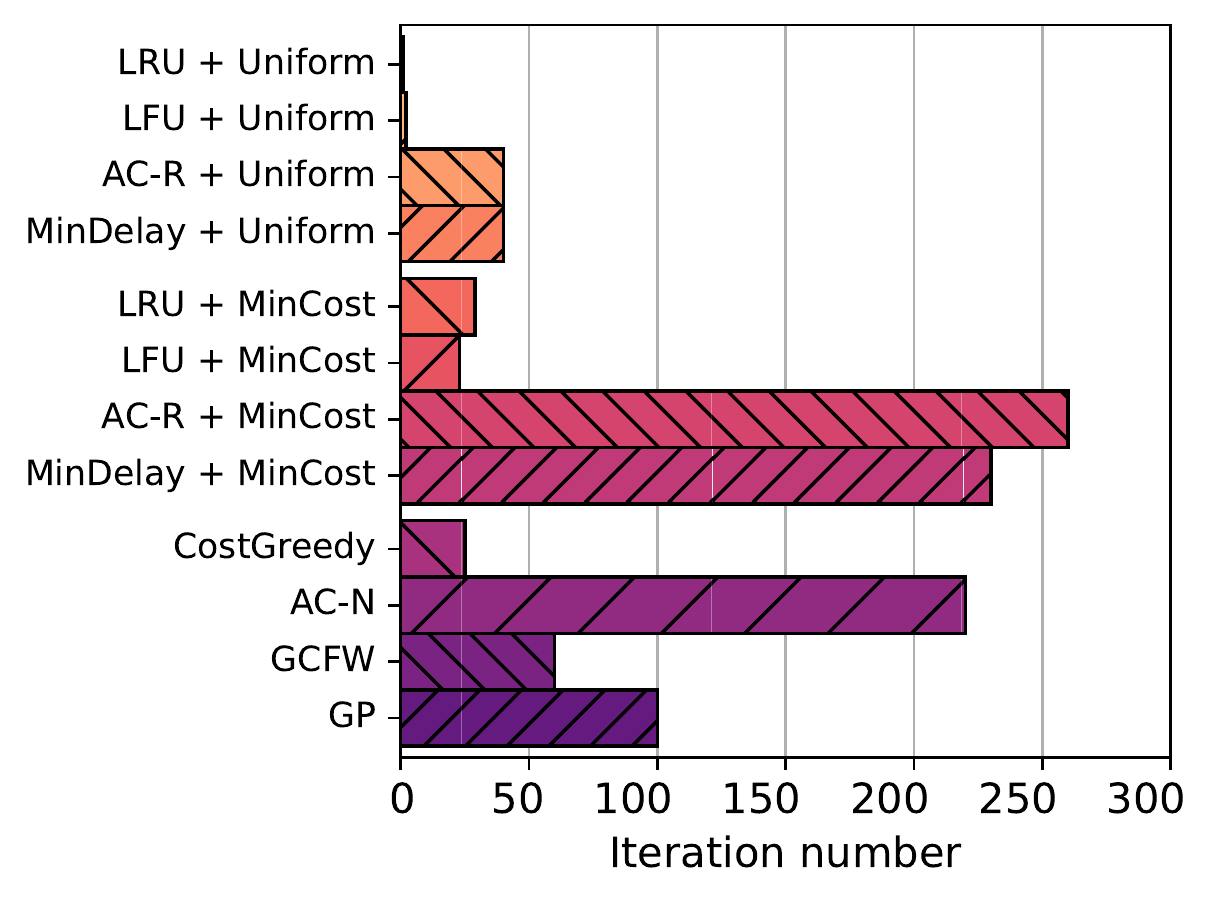}
\caption{Iteration numbers of different methods}
\label{fig_convergenct_iteration_num}
\end{minipage}
\end{figure*}    
\begin{figure*}[htbp]
\centering
\begin{minipage}[t]{0.4\textwidth}
\includegraphics[width=1\textwidth]{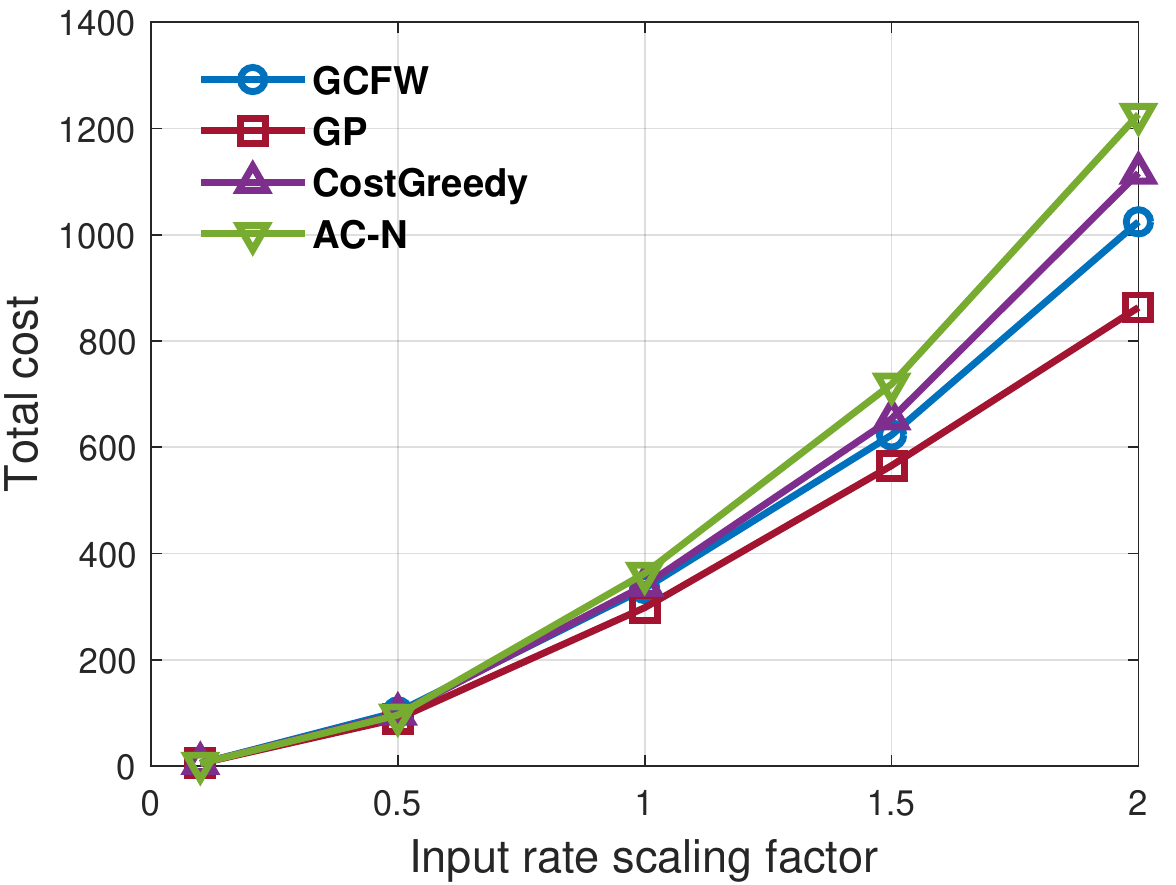}
\caption{Total cost versus universally scaled exogenous request input rates $r_{i}(k)$ in \texttt{grid-25}}
\label{fig_Inputrate}
\end{minipage}
\hspace{5mm}
\begin{minipage}[t]{0.37\textwidth}
\centering
\includegraphics[width=1\textwidth]{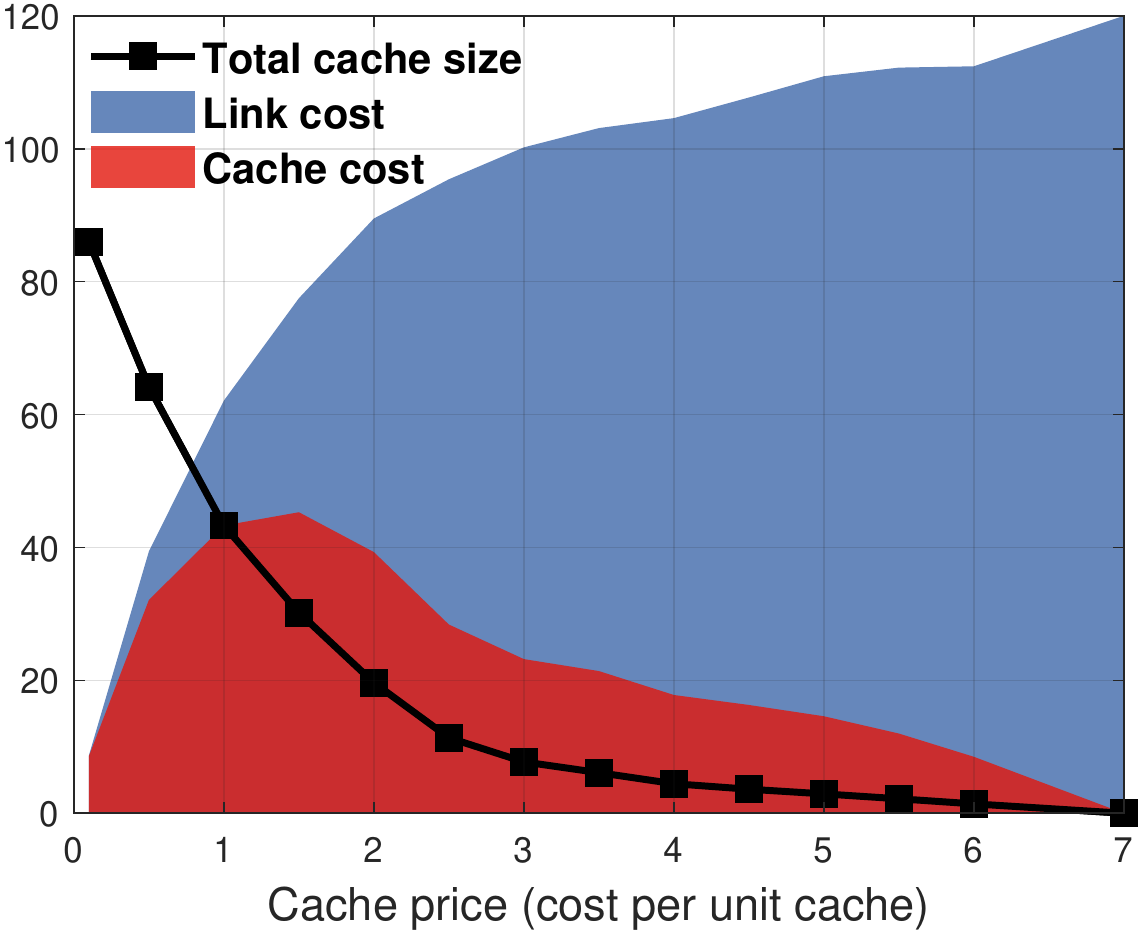}
\caption{ Link costs, cache costs and total cache size versus unit cache cost in \texttt{grid-25} by \texttt{GP}}
\label{fig_price}
\end{minipage}
\end{figure*}

\begin{figure*}[htbp]
\centerline{\includegraphics[width=0.75\textwidth]{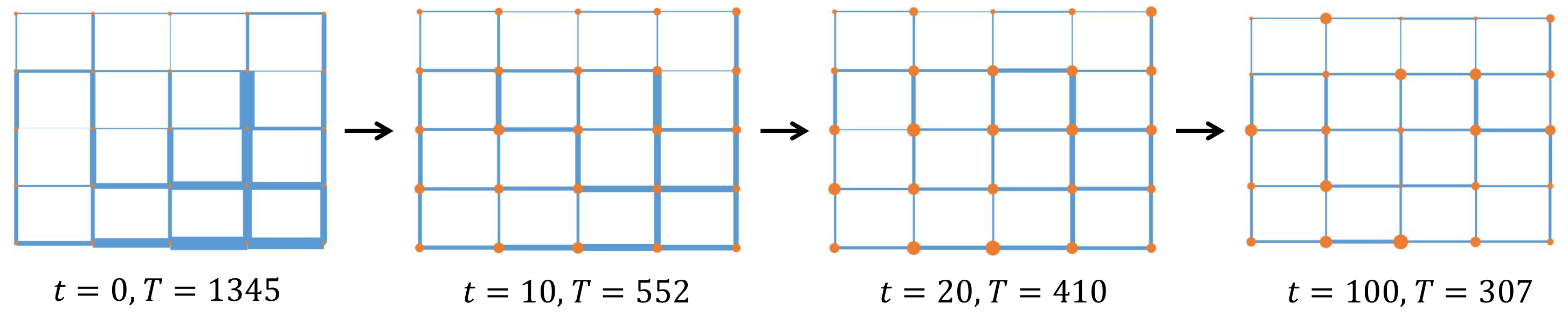}}
\caption{Link flow and cache size evolution in \texttt{grid-25} by \texttt{GP}. Link width and node size are respectively in proportion of link flow and node cache size. $t$ is the period number and $T$ is the measured total cost.}
\label{fig_evolution}
\end{figure*}

We implement proposed \textbf{\texttt{GCFW}} (Algorithm \ref{alg_GCFW}), \textbf{\texttt{GP}} (Algorithm \ref{alg_GP}), and multiple baseline methods summarized in Table \ref{tab:methods} 

\textbf{\texttt{LRU}} and \textbf{\texttt{LFU}} are traditional cache eviction algorithms.
\textbf{\texttt{SP}} (Shortest Path) routes request packets on the reverse of the shortest path from a designated server to the requester.
\textbf{\texttt{AC-R}} (Adaptive Caching with Routing) is a joint routing/caching algorithm proposed by \cite{ioannidis2017jointly}. It uses probabilistic routing among top $k_{\text{SP}} = 3$ shortest paths.
\textbf{\texttt{MinDelay}} is another hop-by-hop joint routing/caching algorithm with convex costs \cite{mahdian2018mindelay}.
It uses Frank-Wolfe algorithm with stepsize $1$ for integer solutions.
\textbf{\texttt{Uniform}} uniformly adds cache capacities by $1$ at all nodes in each period.
\textbf{\texttt{MinCost}} is a heuristic cache deployment algorithm. It adds the cache capacity by $1$ at the node with the highest total cache miss cost in every period\footnote{\texttt{MinCost} can be viewed as a cache miss cost-weighted cache hit maximization. The cache miss cost of a single request $(i,k)$ packet at node $v$ is defined as the sum of $d_{pq}$ at all links $(p,q)$ that the corresponding response travels from generation (either at a cache hit or at the designated server of $k$) till back to node $v$.}. 
\textbf{\texttt{CostGreedy}} is a heuristic joint cache deployment and content placement method. It greedily sets $y_i(k) = 1$ for the node-item pair $(i,k)$ with the largest single-item cache miss cost in every period.
\textbf{\texttt{AC-N}} (Adaptive Caching with Network-wide capacity constraint) is a joint cache deployment and content placement method with a network-wide cache capacity budget \cite{mai2019optimal}. We add the total cache budget by $1$ and re-run \texttt{AC-N} at each period. 

We set $T_{\text{slot} = 10}$ and $L = 20$, and start with the shortest path and $\boldsymbol{y}^0 = \boldsymbol{0}$.
When using \texttt{Uniform} or \texttt{MinCost}, the corresponding content placement 
method is re-run every period to accommodate new cache capacities.
For methods other than \texttt{GCFW} and \texttt{GP}, we run simulation for enough long time and record the lowest period total cost. 
For \texttt{GCFW} and \texttt{GP}, we measure steady state total costs after convergence. 
For \texttt{GCFW}, we set $N = 100$. 
For \texttt{GP}, we use dynamic node blocking, set stepsize $\alpha = 0.01$ and run for $T_{\text{sim}} = 20000$.

\subsection{Results}
We summarize the (normalized) measured total costs in Fig. \ref{fig_bar1}. We divide the methods into three groups.
The first group uses \texttt{Uniform} cache deployment, representing the network performance when the network operator deploys its storage uniformly over the network without optimization.
The second group switches from \texttt{Uniform} to \texttt{MinCost}, representing the performance of heuristic cache deployment optimization methods based on cache utilities. 
The third group represents carefully designed joint optimization methods over cache deployment and content placement.

We observe from Fig. \ref{fig_bar1} that the second group outperforms the first group, and is outperformed by the third group.
This implies that heuristic utility-based cache deployment methods are better than not optimizing, but can be further improved by jointly considering content placement and routing.
Moreover, the proposed online algorithm \texttt{GP} outperforms other methods in all scenarios, with a total cost up to $30\%$ less than the second best algorithm -- the proposed \texttt{GCFW} combined with \texttt{SP}. 
Specifically, the improvement of \texttt{GP} against \texttt{GCFW} is more significant in scenarios with more routing choices (e.g., \texttt{grid-100} and \texttt{DTelekom}), and diminishes when routing choice is limited (e.g., \texttt{full-tree}). 
We credit such performance improvement to three fundamental advantages in our model: the congestion-dependent link cost functions, the hop-by-hop routing scheme, and the unified modeling of link and cache costs.

\label{sec:result_grid-25}
To help better understand the behavior of proposed algorithms, we present more refined experiments on the scenario \texttt{grid-25}.

\noindent\textbf{Convergence.} 
We plot in Fig. \ref{fig_convergence} the convergence trajectory of measured and theoretical total cost by \texttt{GCFW}+\texttt{SP} and \texttt{GP} in \texttt{grid-25}.
The measured cost refers to the actual link costs calculated from real link flows measured by number of response packets, and the actual cache costs calculated from real cache sizes after rounding.
The theoretical cost refers to the flow-level cost $T$ given by \eqref{Objective_cache}, calculated using the pre-given input rates $[r_{i}(k)]$ and the relaxed variable $(\boldsymbol{\phi}^t, \boldsymbol{y}^t)$.
We can see from Fig. \ref{fig_convergence} that, with the continuous relaxation, randomized packet forwarding and randomized rounding, our theoretical model accurately reflects the real network behavior.

\noindent\textbf{Runtime.}
We compare the total iteration numbers across different methods in Fig. \ref{fig_convergenct_iteration_num}.
The iteration number of \texttt{CGFW} and \texttt{GP} is the number of periods until convergence.
The iteration number of other methods is the period number before reaching the minimum total cost multiplied by the algorithm iteration number within each period.
For \texttt{LRU}, \texttt{LFU} and \texttt{CostGreedy}, one iteration is counted in each period.
We assume \texttt{AC-R}, \texttt{MinDelay} and \texttt{AC-N} iterate every slot, yielding $L$ iterations per period.
Compared with double-loop methods (i.e., an outer-loop for cache deployment, and an inner-loop for content placement and routing), \texttt{GCFW} and \texttt{GP} use fewer iterations since they have only one loop, which jointly determines the cache deployment and content placement.     

\noindent\textbf{Congestion mitigating.}
Since we model non-linear congestion-dependent link costs, the ability of mitigating network congestion is expected to be an important feature of the proposed algorithms. 

Fig. \ref{fig_evolution} graphically illustrates the evolution of link flows and cache sizes across the network as \texttt{GP} goes on.
The requests and designated servers are randomly generated according to our previous assumptions in Section \ref{sec:simulator setting} and Table \ref{tab:scenarios}.
We observe that severe link congestion is gradually mitigated by properly tuning routing and caching strategies.
Fig. \ref{fig_Inputrate} shows the total cost of \texttt{grid-25} by methods in the third group (\texttt{CostGreedy}, \texttt{AC-N}, \texttt{GCFW}, \texttt{GP}), where other parameters remain unchanged except all exogenous request input rates $r_i(k)$ are scaled by a same factor. 
The network congestion becomes more severe as the scaling factor growing, and we observe that \texttt{GP} is the most resilient method to such congestion.

\noindent\textbf{Routing-caching tradeoff.}
As a fundamental motivation of this paper, we investigate the tradeoff between routing cost and cache deployment cost in \texttt{grid-25}.
We set the link cost to be linear and plot in Fig. \ref{fig_price} the optimized link and cache costs as well as the corresponding total cache size against different unit cache cost $b_i$.
We observe that, with very high unit cache cost, no cache is deployed.
As the unit cache cost dropping, the total cache size increases, the total cost decreases, and the cache cost takes a gradually more significant portion of the total cost.


\section{Future directions}
In this paper, we mainly focused on social welfare maximization (i.e., cost minimization) from the perspective of cooperative nodes.
Nevertheless, our formulation and framework provide insights beyond the scope of this particular problem. 
We next give some potential future directions.

\label{sec:extensions}
\noindent\textbf{Congestion control and fairness.}
The network may not be capable of fully handling all requests. It can choose to admit only a fraction of request packets. A concave utility function of the actual admitted rate can be assigned on each request to achieve network congestion control functionality.
Such concave user-associated utilities can also address the inter-user fairness issue, e.g., \cite{liu2021fair}.

\noindent\textbf{Non-cooperative nodes.}
Instead of maximizing social welfare among cooperative nodes, in a practical network, it is possible that each node (or group of nodes) is selfish and seeks to minimize its own cost.
Investigating the game-theoretical behavior of non-cooperative nodes in an elastic cache network may become a future direction.

\noindent\textbf{Optimal cache pricing.}
In this paper, the cache deployment cost (i.e., cache price) is given and fixed. However, from the perspective of the cache provider, the question of how to price its cache service is also intriguing. 
Studying the cache provider's optimal pricing strategy to maximize its income  (i.e. the total cache cost) from a rational network operator is also a related open problem.
    

\section{Conclusion}
\label{sec:conclusion}
We aim to minimize the sum of routing cost and cache deployment cost by jointly determining cache deployment, content placement and routing strategies, in a network with arbitrary topology and general convex costs.
In the fixed-routing special case, we show that the objective is a \emph{DR-submodular + concave} function, and propose a Gradient-combining Frank-Wolfe algorithm with $(\frac{1}{2},1)$ approximation.
For the general case, we propose the KKT condition and a modification. 
The modified condition suggests each node handles arrival requests in the way that achieves minimum marginal cost.
We propose a distributed and adaptive online algorithm for the general case that converges to the modified condition. 
We demonstrate in simulation that our proposed algorithms significantly outperform baseline methods in multiple network scenarios.

\bibliographystyle{ACM-Reference-Format}
\bibliography{sample-base}

\appendix

\section{Proof of Proposition \ref{Prop_cache_NPhard}}
\label{prof_prop_NPhard}
With fixed cache capacities, fixed routing paths and linear link costs, problem \eqref{Objective_cache} is reduced to the continuous-relaxed offline caching problem (problem (8) in \cite{ioannidis2018adaptive}). 
We next show  problem (8) in \cite{ioannidis2018adaptive} is NP-hard.

It is shown in \cite{ioannidis2018adaptive} that, for any given continuous solution to problem (8) in \cite{ioannidis2018adaptive}, \texttt{pipage rounding} can always round it to an integer solution with no-worse objective value in polynomial time.
Therefore, suppose that problem (8) in \cite{ioannidis2018adaptive} is not NP-hard (i.e., if problem (8) in \cite{ioannidis2018adaptive} can be solved in polynomial time), then the corresponding integer caching problem (problem (4) in \cite{ioannidis2018adaptive}) can also be solved in polynomial time (by solving its continuous-relaxed version and rounding).

However, this contradicts with the fact that problem (4) in \cite{ioannidis2018adaptive} is NP-complete \cite{dehghan2015complexity}.
Therefore, problem (8) in \cite{ioannidis2018adaptive} must be NP-hard, thus our problem \eqref{Objective_cache} is also NP-hard.

\section{Proof of Lemma \ref{lemma:fixroute:submodular}}
\label{proof_lemma_submodular}
The non-negativity and monotonicity of $A(\boldsymbol{y})$ as well as the convexity of $B(\boldsymbol{y})$ are obvious.
We prove the DR-submodularity of $A(\boldsymbol{y})$ by showing that\footnote{This criteria can be found in \cite{bian2017guaranteed}}
\begin{equation}
    \frac{\partial^2 A(\boldsymbol{y})}{\partial y_{i_1}(k_1) \partial y_{i_2}(k_2) } \leq 0, \quad \forall i_1,i_2 \in \mathcal{V}, k_1, k_2 \in \mathcal{C}.
    \label{fixroute:proof_submod_goal}
\end{equation}
By the definition of $A(\boldsymbol{y})$, for any $i_1$ and $k_1$, we have
\begin{equation*}
\begin{aligned}
    \frac{\partial A(\boldsymbol{y})}{\partial y_{i_1}(k_1)} &= - \sum_{(i,j) \in \mathcal{E}} \frac{\partial D_{ij}(F_{ij})}{\partial y_{i_1}(k_1)}  
    = -  \sum_{(i,j)\in \mathcal{E}}D^\prime_{ij}(F_{ij})\frac{\partial F_{ij}}{\partial y_{i_1}(k_1)}
\end{aligned}
\end{equation*}
and thus for any $i_2$ and $k_2$, we have
\begin{equation}
\begin{aligned}
&\frac{\partial^2 A(\boldsymbol{y})}{\partial y_{i_1}(k_1) \partial y_{i_2}(k_2) }
\\&= -   \sum_{(i,j)\in\mathcal{E}} \left( \frac{\partial D^\prime_{ij}(F_{ij})}{\partial y_{i_2}(k_2)} \frac{\partial F_{ij}}{\partial y_{i_1}(k_1)} 
+ D^\prime_{ij}(F_{ij}) \frac{ \partial^2 F_{ij}}{\partial y_{i_1}(k_1) \partial y_{i_2}(k_2)} \right)
\\ &= -  \sum_{(i,j)\in\mathcal{E}} \left( D^{\prime\prime}_{ij}(F_{ij}) \frac{\partial F_{ij}}{\partial y_{i_2}(k_2)} \frac{\partial F_{ij}}{\partial y_{i_1}(k_1)} + D^\prime_{ij}(F_{ij}) \frac{ \partial^2 F_{ij}}{\partial y_{i_1}(k_1) \partial y_{i_2}(k_2)} \right) 
\end{aligned}
\label{fixroute:proof_submod_secderi}
\end{equation}
By the assumption that $D_{ij}(\cdot)$ is increasing convex, we know that
\begin{equation*}
    D^\prime_{ij}(F_{ij}) \geq 0, \quad D^{\prime\prime}_{ij}(F_{ij}) \geq 0.
\end{equation*}
Meanwhile, by \eqref{fixroute:Fij} we have
\begin{equation*}
    \frac{\partial F_{ij}}{\partial y_{i_1}(k_1)} \leq 0, \quad \frac{\partial F_{ij}}{\partial y_{i_2}(k_2)} \leq 0, \quad \quad \frac{\partial^2 F_{ij}}{\partial y_{i_1}(k_1)\partial y_{i_2}(k_2)} \geq 0.
\end{equation*}
Therefore by \eqref{fixroute:proof_submod_secderi} we know \eqref{fixroute:proof_submod_goal} holds and thus $A(\boldsymbol{y})$ is submodular, which completes the proof.

\section{Loss of DR-submodularity in joint routing and caching}
\label{appendix_non_DRsubmodular}

\begin{figure*}[t!]
    \centering
    \subcaptionbox{Without alternative path}{\includegraphics[width = 0.37\linewidth]{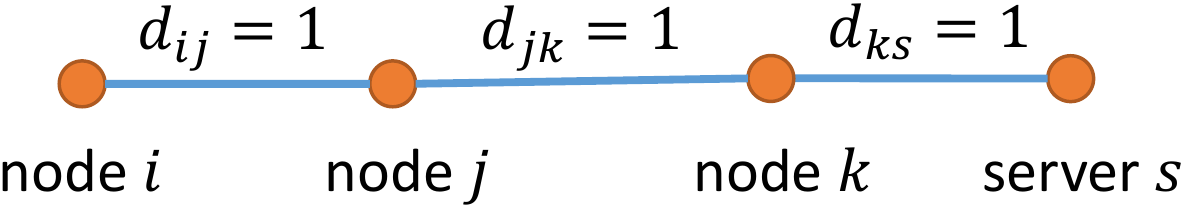}}
    \hspace{5mm}
      \subcaptionbox{With alternative path}{\includegraphics[width = 0.37\linewidth]{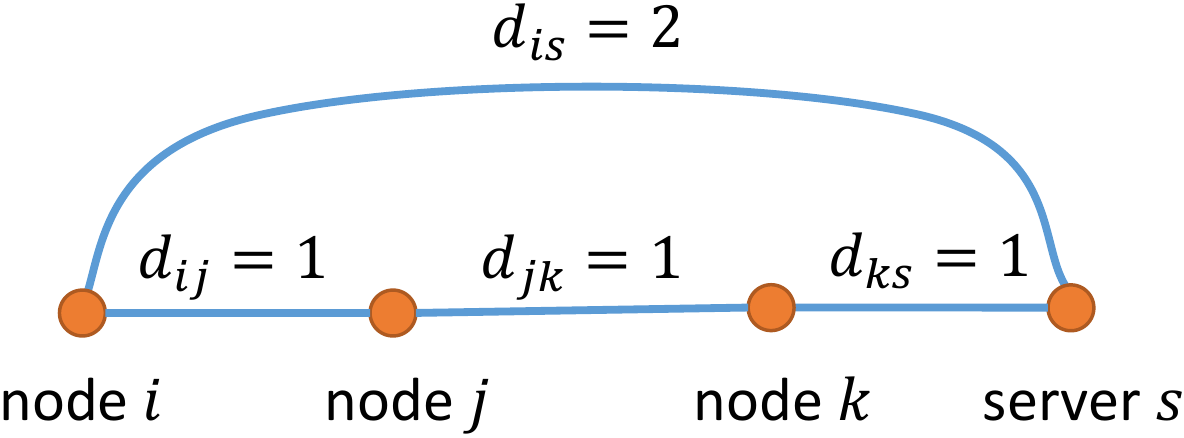}}
    \caption{The routing cost loses its DR-supermodularity, if requests are allowed to choose between paths. }
    \label{fig:non_submodular}
\end{figure*}

We demonstrate the basic idea of DR-submodular and the loss of DR-submodularity in the general case of joint routing and caching by the example in Fig. \ref{fig:non_submodular}.
For simplicity, in Fig. \ref{fig:non_submodular}, we assume there is only one item, and omit the item notation.
We use a linear cost $D_{vu}(F_{vu}) = d_{vu}f_{vu}$, and assume that $d_{uv} = d_{vu}$, for all links $(v,u)$.
Moreover, we assume in the network, only node $i$ makes request, with rate $r_i = 1$, where the designated server is located at $s$.

As an analog to the DR-submodular gain $A(\boldsymbol{y})$ in the fixed-routing case \eqref{fixroute:Obj}, and since the optimal routing path with linear link costs can be efficiently found, we define the \emph{routing-caching gain} as 
$$G(\boldsymbol{y}) = T_{\min}(\boldsymbol{0}) - T_{\min}(\boldsymbol{y}),$$
where $T_{\min}(\boldsymbol{y})$ is the optimal total routing cost given caching strategy $\boldsymbol{y}$.
Note that such ``optimal'' is in the sense that requests choose the minimum-cost paths.

For example, in the network given by Fig. \ref{fig:non_submodular} (a), we have $T_{\min}^{\text{(a)}}(\boldsymbol{0}) = 3$ as there is only one available path $i \to j \to k \to s$.
In Fig. \ref{fig:non_submodular} (b), we have $T_{\min}^{\text{(b)}}(\boldsymbol{0}) = 2$, since when all three nodes $i$, $j$ and $k$ do not cache the item, the minimum-cost routing path is $i \to s$.
For simplicity, we use ``$\left( y_{v} \text{ for all } y_{v} \neq 0 \right)$'' to denote the caching strategy $\boldsymbol{y}$ such that all elements other than $y_v$ are zero. For example, we have $T_{\min}^{\text{(a)}}\left(y_j = 0.5\right) = 2$, since the minimum routing cost in (a) equals $2$ if only $j$ has non-zero caching variable with $y_j = 0.5$. 
Similarly, $T_{\min}^{\text{(b)}}\left(y_j = 0.5\right) = 2$.
Therefore, $G^{\text{(a)}}(y_j = 0.5) = 1$ and $G^{\text{(b)}}(y_j = 0.5) = 0$.

Before introducing the DR-submodularity in (a) and non-DR-submodularity in (b), we first give the concept of \emph{cache operation} and \emph{diminishing return} (similar concepts can be found in various DR-submodular papers, e.g., \cite{bian2019optimal}).
For a cache strategy $\boldsymbol{y} \in [\boldsymbol{0},\boldsymbol{1}]$ and an incremental amount ${\Delta}\boldsymbol{y} \geq \boldsymbol{0}$ such that $\boldsymbol{y} + {\Delta}\boldsymbol{y} \in [\boldsymbol{0},\boldsymbol{1}]$, we denote by ${\Delta}\boldsymbol{y} \big| \boldsymbol{y}$ the operation that increasing the strategy from $\boldsymbol{y}$ to $\boldsymbol{y} + {\Delta}\boldsymbol{y}$ (we say an operation ${\Delta}\boldsymbol{y} \big| \boldsymbol{y}$ is valid if it satisfies $\boldsymbol{y} \in [\boldsymbol{0},\boldsymbol{1}]$,  ${\Delta}\boldsymbol{y} \geq \boldsymbol{0}$ and $\boldsymbol{y} + {\Delta}\boldsymbol{y} \in [\boldsymbol{0},\boldsymbol{1}]$), and denote by $g({\Delta}\boldsymbol{y} \big| \boldsymbol{y})$ the \emph{return} by such operation, i.e.,
\begin{equation*}
    g({\Delta}\boldsymbol{y} \big| \boldsymbol{y}) = G(\boldsymbol{y} + \boldsymbol{\Delta}\boldsymbol{y}) - G(\boldsymbol{y}).
\end{equation*}

With such setting, if for any $\boldsymbol{y}_1$, $\boldsymbol{y}_2$ and $\Delta\boldsymbol{y}$, such that $\boldsymbol{y}_1 \leq \boldsymbol{y}_2$, and operations ${\Delta}\boldsymbol{y} \big| \boldsymbol{y}_1$, ${\Delta}\boldsymbol{y} \big| \boldsymbol{y}_2$ are both valid, it holds that $g(\boldsymbol{\Delta}\boldsymbol{y} \big| \boldsymbol{y}_1) \geq g(\boldsymbol{\Delta}\boldsymbol{y} \big| \boldsymbol{y}_2)$,
we say the gain $G(\cdot)$ has the \emph{diminishing-return} property (i.e., DR-submodular).
Namely, in the caching context, DR-submodularity refers to the following intuition: The more you already cache, the less improvement you can get by caching new things.

It is shown that in the network in Fig. \ref{fig:non_submodular} (a), the gain $G^{\text{(a)}}(\boldsymbol{y})$ is DR-submodular (see Lemma \ref{lemma:fixroute:submodular}).
For example, let $\boldsymbol{y}_1$ be $\boldsymbol{0}$, $\boldsymbol{y}_2$ be $(y_j = 0.5)$ and $\Delta\boldsymbol{y}$ be $(y_k = 1)$, we have $g^{\text{(a)}}({\Delta}\boldsymbol{y} \big| \boldsymbol{y}_1) = 1$, and $g^{\text{(a)}}({\Delta}\boldsymbol{y} \big| \boldsymbol{y}_2) = 0.5$, that is, for the same operation ``cache at $k$'', the improvement in caching gain is decreased by $0.5$ if we already have $y_j = 0.5$.

However, such DR-submodularity does not hold in Fig. \ref{fig:non_submodular} (b), where an alternative path exists directly from $i$ to $s$.
To see this, we still let $\boldsymbol{y}_1$ be $\boldsymbol{0}$, $\boldsymbol{y}_2$ be $(y_j = 0.5)$ and $\Delta\boldsymbol{y}$ be $(y_k = 1)$.
It is easy to see $g^{\text{(b)}}({\Delta}\boldsymbol{y} \big| \boldsymbol{y}_1) = 0$, since no matter $k$ caches the item or not, the minimum routing cost is $2$.
But one can see that $g^{\text{(b)}}({\Delta}\boldsymbol{y} \big| \boldsymbol{y}_2) = 0.5$, since $G^{\text{(b)}}(y_j = 0.5) = 0$ and $G^{\text{(b)}}(y_j = 0.5, y_k = 1) = 0.5$.
That is, if we already have $y_j = 0.5$, additionally caching at $k$ will generate higher return than if not.

We remark that, intuitively speaking, such non-DR-submodularity is because when increasing $\boldsymbol{y}$, the optimal routing path may change, and consequently ``activating'' some nodes that were not on the previous optimal path (i.e., increasing the flow rate on these nodes). Therefore, additionally caching on these ``activated'' nodes will generates more improvements than previously.

Ioannidis et al. \cite{ioannidis2017jointly} devised a technique unifying routing and caching variables, and a caching gain can be shown DR-submodular in $y_{ik}$ and $(1-\phi_{ij}(k))$. 
However, the caching gain is defined w.r.t. an upper bound given by $T(\boldsymbol{1},\boldsymbol{0})$, i.e., the total cost when (1) no cache is deployed, and (2) nodes duplicate and broadcast every arrival request to all neighbors.
We do not use this technique since in our context, such upper bound is not likely to be finite.

\section{Proof of Theorem \ref{thm_necessary_cache}}
\label{proof_thm_necessary}
Following \cite{gallager1977minimum}, we write the Lagrangian function of \eqref{Objective_cache} as

\begin{equation*}
\begin{aligned}
    L(\boldsymbol{\phi},\boldsymbol{y},\boldsymbol{\lambda},\boldsymbol{\mu}) 
    &= T(\boldsymbol{\phi},\boldsymbol{y}) - \sum_{i \in \mathcal{V}} \sum_{k \in \mathcal{C}} \lambda_{ik}\left( y_{i}(k) + \sum_{j \in \mathcal{V}}\phi_{ij}(k) - \mathbbm{1}_{i \in \mathcal{S}_k} \right)    
    \\&- \sum_{i \in \mathcal{V}} \sum_{k \in \mathcal{C}} \left(\mu_{i0k}y_{i}(k) + \sum_{j \in \mathcal{V}}\mu_{ijk}\phi_{ij}(k) \right)
\end{aligned}
\end{equation*}
where $\boldsymbol{\lambda} = [\lambda_{ik}]_{i \in \mathcal{V}, k \in \mathcal{C}} \in \mathbb{R}^{|\mathcal{V}||\mathcal{C}|}$, $\boldsymbol{\mu} = [\mu_{ijk}]_{i \in \mathcal{V}, j \in \left\{ 0 \cup \mathcal{V}\right\}, k \in \mathcal{C}} \in \left(\mathbb{R}^+\right)^{|\mathcal{V}|(|\mathcal{V}|+1)|\mathcal{C}|}$ are the Lagrangian multiplier vectors, with the complementary slackness holds, i.e.,
\begin{equation*}
\begin{aligned}
    & \mu_{i0k}y_{i}(k) = 0, \quad \forall i \in \mathcal{V}, k \in \mathcal{C}.
    \\&\mu_{ijk}\phi_{ij}(k) = 0, \quad \forall i,j \in \mathcal{V},  k \in \mathcal{C},
\end{aligned}
\end{equation*}

Taking the derivative of $L$ w.r.t. $\boldsymbol{y}$ and $\boldsymbol{\phi}$ and set to $0$, we have
\begin{equation*}
\begin{aligned}
    \frac{\partial T}{\partial y_{i}(k)} = \lambda_{ik} + \mu_{i0k}, \quad \frac{\partial T}{\partial \phi_{ij}(k)} = \lambda_{ik} + \mu_{ijk}.
\end{aligned}
\end{equation*}
Combining with the complementary slackness, we arrive at the KKT necessary condition.

\section{Proof of Theorem \ref{thm_BoundedGap_cache}}
\label{proof_bounded_gap}
We start with the link cost.
Note that by the condition in Theorem \ref{thm_BoundedGap_cache}, for link $(i,j)$ with $\phi_{ij}(k) > 0$,
\begin{equation}
    D^\prime_{ji}(F_{ji}) + \frac{\partial T}{\partial r_j(k)} = \delta_{i}(k).
    \label{Proof_cache_1}
\end{equation}
Multiply \eqref{Proof_cache_1} by $\phi_{ij}(k)$ and sum over all such $j$, we have that for any $i \in \mathcal{V}$ and $k \in \mathcal{C}$,
\begin{equation*}
    \sum_{j: \phi_{ij}(k) > 0} \phi_{ij}(k)\left(D^\prime_{ji}(F_{ji}) + \frac{\partial T}{\partial r_j(k)}\right) = \delta_{i}(k) \sum_{j: \phi_{ij}(k) > 0}\phi_{ij}(k).
\end{equation*}
We let $p_{ik} = \sum_{j\in\mathcal{N}(j)}\phi_{ij}(k)$ and combining with \eqref{pT_pr_cache}, the above is equivalent to
\begin{equation}
    \frac{\partial T}{\partial r_i(k)} = p_{ik}\delta_{i}(k).
\label{Proof_cache_1.5}
\end{equation}
We repeat the condition for routing variable in Theorem \ref{thm_BoundedGap_cache} as: for all $j \in \mathcal{N}(i)$,
\begin{equation}
    D^\prime_{ji}(F_{ji}) + \frac{\partial T}{\partial r_j(k)} \geq \delta_{i}(k).
\label{Proof_cache_2}
\end{equation}

We now bring in the arbitrarily chosen feasible $(\boldsymbol{\phi}^\dagger,\boldsymbol{y}^\dagger)$. Multiply \eqref{Proof_cache_2} by $\phi_{ij}^\dagger(k)$ and sum over all $j \in \mathcal{N}(i)$, we have
\begin{equation*}
    \sum_{j \in \mathcal{N}(i)} \left(D^\prime_{ji}(F_{ji}) + \frac{\partial T}{\partial r_j(k)}\right)\phi_{ij}^\dagger(k) \geq \delta_{i}(k)\left(\sum_{j \in \mathcal{N}(i)}\phi_{ij}^\dagger(k)\right).
\end{equation*}
Rearrange the term and note that $p_{ik}^\dagger = \sum_{j \in \mathcal{N}(i)}\phi_{ij}^\dagger(k)$, we have
\begin{equation}
    \sum_{j \in \mathcal{N}(i)} D^\prime_{ji}(F_{ji})\phi_{ij}^\dagger(k)
    \geq p_{ik}^\dagger \delta_{i}(k) - \sum_{j \in \mathcal{N}(i)}\phi_{ij}^\dagger(k) \frac{\partial T}{\partial r_j(k)}
\label{Proof_cache_3}
\end{equation}

Multiply \eqref{Proof_cache_3} by $t_{i}^\dagger(k)$ and note that $f_{ji}^\dagger(k) = t_{i}^\dagger(k)\phi_{ij}^\dagger(k)$, we have
\begin{equation}
    \sum_{j \in \mathcal{N}(i)}D^\prime_{ji}(F_{ji})f_{ji}^\dagger(k) \geq p_{ik}^\dagger t_{i}^\dagger(k)\delta_{i}(k)
    -  \sum_{j \in \mathcal{N}(i)}\phi_{ij}^\dagger(k)t_i^\dagger(k)\frac{\partial T}{\partial r_j(k)}
\label{Proof_cache_4}
\end{equation}

Sum \eqref{Proof_cache_4} over all $i \in \mathcal{V}$ and $k \in \mathcal{C}$, we have
\begin{equation}
\begin{aligned}
&\sum_{i \in \mathcal{V}}\sum_{j\in\mathcal{N}(i)} D^\prime_{ji}(F_{ji}) F_{ji}^\dagger \geq \sum_{i\in\mathcal{V}}\sum_{k \in \mathcal{C}}p_{ik}^\dagger t_i^\dagger(k)\delta_{i}(k) 
\\&- \sum_{i \in \mathcal{V}}\sum_{j \in \mathcal{N}(i)}\sum_{k \in \mathcal{C}}\phi_{ij}^\dagger(k)t_i^\dagger(k)\frac{\partial T}{\partial r_j(k)}.
\label{Proof_cache_5}
\end{aligned}
\end{equation}

Recall that $\sum_{i \in \mathcal{N}(j)}\phi_{ij}^\dagger(k) t_i^\dagger(k) = t_j^\dagger(k) - r_j(k)$, we swap the summation order in the last term of \eqref{Proof_cache_5}, and replace it by
\begin{equation*}
\begin{aligned}
    &- \sum_{j \in \mathcal{V}}\sum_{k \in \mathcal{C}}\frac{\partial T}{\partial r_j(k)}\left(\sum_{i \in \mathcal{N}(j)}\phi_{ij}^\dagger(k)t_i^\dagger(k)\right)
    \\&= \sum_{i \in \mathcal{V}}\sum_{k \in \mathcal{C}}\frac{\partial T}{\partial r_i(k)}\left(r_i(k) - t_i^\dagger(k)\right).
\end{aligned}
\end{equation*}

Therefore, \eqref{Proof_cache_5} is equivalent to
\begin{equation*}
\begin{aligned}
    &\sum_{i \in \mathcal{V}}\sum_{j\in\mathcal{N}(i)} D^\prime_{ji}(F_{ji}) F_{ji}^\dagger \geq
    \sum_{i\in\mathcal{V}}\sum_{k \in \mathcal{C}}p_{ik}^\dagger t_i^\dagger(k)\delta_{i}(k) 
    \\&+ \sum_{i \in \mathcal{V}}\sum_{k \in \mathcal{C}}\frac{\partial T}{\partial r_i(k)}\left(r_i(k) - t_i^\dagger(k)\right).
\end{aligned}
\end{equation*}

Recall \eqref{Proof_cache_1.5} and replace $\partial T/ \partial r_i(k)$, the above is equivalent to
\begin{equation}
\begin{aligned}
    &\sum_{i \in \mathcal{V}}\sum_{j\in\mathcal{N}(i)} D^\prime_{ji}(F_{ji}) F_{ji}^\dagger \geq
    \sum_{i\in\mathcal{V}}\sum_{k \in \mathcal{C}} \left(p_{ik}^\dagger - p_{ik}\right)t_i^\dagger(k)\delta_{i}(k) 
    \\& + \sum_{i \in \mathcal{V}}\sum_{k \in \mathcal{C}}\frac{\partial T}{\partial r_i(k)}r_i(k) .
\end{aligned}
\label{Proof_cache_6}
\end{equation}


On the other hand, we next give an analog of \eqref{Proof_cache_6} when $\boldsymbol{\phi}^\dagger = \boldsymbol{\phi}$.
Notice that \eqref{pT_pr_cache} is equivalent to
\begin{equation*}
    \sum_{j \in \mathcal{N}(i)} D^\prime_{ji}(F_{ji}) \phi_{ij}(k)= \frac{\partial T}{\partial r_i(k)} - \sum_{j \in \mathcal{N}(i)}\phi_{ij}(k)\frac{\partial T}{\partial r_j(k)}.
\end{equation*}
Multiply the above by $t_i(k)$ and sum over $i\in\mathcal{V}$ and $k \in \mathcal{C}$, we have
\begin{equation*}
\begin{aligned}
    &\sum_{i \in \mathcal{V}} \sum_{j \in \mathcal{N}(i)} D^\prime_{ji}(F_{ji}) F_{ji} = \sum_{i \in \mathcal{V}} \sum_{k \in \mathcal{C}}t_i(k) \frac{\partial T}{\partial r_i(k)} 
    \\&- \sum_{i \in \mathcal{V}} \sum_{k \in \mathcal{C}} \sum_{j \in \mathcal{N}(i)} \phi_{ij}(k)t_i(k)\frac{\partial T}{\partial r_j(k)}.
\end{aligned}
\end{equation*}
Replace the last term in above with $\sum_{i}\sum_{k} \left(r_i(k) - t_i(k)\right)\partial T / \partial r_i(k)$, 
we have
\begin{equation}
    \sum_{i \in \mathcal{V}} \sum_{j \in \mathcal{N}(i)} D^\prime_{ji}(F_{ji}) F_{ji} =
     \sum_{i \in \mathcal{V}} \sum_{k \in \mathcal{C}} \frac{\partial T}{\partial r_i(k)}r_i(k).
\label{Proof_cache_7}
\end{equation}

Subtract \eqref{Proof_cache_7} from \eqref{Proof_cache_6}, we have 
\begin{equation}
\sum_{(j,i) \in \mathcal{E}} D^\prime_{ji}(F_{ji}) \left(F_{ji}^\dagger - F_{ji}\right)
\geq  \sum_{i\in\mathcal{V}}\sum_{k \in \mathcal{C}} \left(p_{ik}^\dagger - p_{ik}\right)t_i^\dagger(k)\delta_{i}(k) 
\label{Proof_cache_8}
\end{equation}

Next, we consider the caching cost. Recall that by the condition in Theorem \ref{thm_BoundedGap_cache}, we have
\begin{equation*}
     B^\prime_{i}(Y_{i}) \geq t_i(k)\delta_{i}(k), 
\end{equation*}
and the equality holds when $y_{i}(k) > 0$.
Therefore, 
\begin{equation}
\begin{aligned}
        \sum_{i\in\mathcal{V}} B^\prime_{i}(Y_{i})\left(Y_{i}^\dagger - Y_{i}\right) 
    &=  \sum_{i\in\mathcal{V}} \sum_{k \in \mathcal{C}} B^\prime_{i}(Y_{i})\left(y_{i}^\dagger(k) - y_{i}(k)\right) 
    \\& \geq \sum_{i \in \mathcal{V}} \sum_{k \in \mathcal{C}}t_i(k)\delta_{i}(k)\left(y_{i}^\dagger(k) - y_{i}(k)\right)
\end{aligned}
\label{Proof_cache_9}
\end{equation}
Note that in the RHS of \eqref{Proof_cache_9}, for the case $i \not \in \mathcal{S}_k$, we must have $y_{i}^\dagger(k) = 1- p_{ik}^\dagger$ and $y_{i}(k) = 1- p_{ik}$. For the case $i \in \mathcal{S}_k$, we have $p_{ik}^\dagger = p_{ik} \equiv 0$ as well as $y_{i}^\dagger(k) = y_{i}(k) \equiv 0$.
Thus \eqref{Proof_cache_9} implies that
\begin{equation}
\begin{aligned}
        \sum_{i\in\mathcal{V}} B^\prime_{i}(Y_{i})\left(Y_{i}^\dagger - Y_{i}\right) 
    \geq \sum_{i\in\mathcal{V}} \sum_{k \in \mathcal{C}}t_i(k)\delta_{i}(k)\left(p_{ik} - p_{ik}^\dagger\right).
\end{aligned}
\label{Proof_cache_9.5}
\end{equation}

Finally we compare $T = (\boldsymbol{\phi},\boldsymbol{y})$ and $T^\dagger = (\boldsymbol{\phi}^\dagger,\boldsymbol{y}^\dagger)$.
Note that $T$ as a function is jointly convex in the total link rates $\boldsymbol{F} = [F_{ij}]_{(i,j)\in\mathcal{E}}$ and the occupied cache sizes $\boldsymbol{Y} = [Y_{i}]_{i\in\mathcal{V}}$ due to the convexity of $D_{ij}(\cdot)$ and $B_{m}(\cdot)$.
Thus we have
\begin{equation}
\begin{aligned}
    T^\dagger - T & \geq \left(\boldsymbol{F}^\dagger - \boldsymbol{F} \right) \nabla_{\boldsymbol{F}}T + \left(\boldsymbol{Y}^\dagger - \boldsymbol{Y} \right) \nabla_{\boldsymbol{Y}}T
    \\ &= \sum_{(i,j) \in \mathcal{E}}\left(F_{ij}^\dagger - F_{ij}\right)D^\prime_{ij}(F_{ij})
    + \sum_{i \in \mathcal{V}} \left(Y_{i}^\dagger - Y_{i}\right)B^\prime_{i}(Y_{i}).
\end{aligned}
\label{Proof_cache_10}
\end{equation}

Substituting \eqref{Proof_cache_8} and \eqref{Proof_cache_9.5} into \eqref{Proof_cache_10}, we have
\begin{equation*}
    T^\dagger - T \geq \sum_{i \in \mathcal{V}}\sum_{k \in \mathcal{C}}\delta_{i}(k)\left(p_{ik}^\dagger - p_{ik}\right)\left(t_{i}(k)^\dagger - t_{i}(k)\right)
\end{equation*}
which completes the proof.

We remark that function $T(\boldsymbol{\phi},\boldsymbol{y})$ is a summation of a convex function and a geodesic convex function. 
Specifically, the caching cost part $\sum_{i \in \mathcal{V}} B_i(Y_i)$ is dependent only on caching strategies $\boldsymbol{Y}$, with the convexity obvious due to the convex assumption of $B_i$ functions.
The routing cost part $\sum_{(i,j) \in \mathcal{E}} D_{ij}(F_{ij})$ is dependent only on routing strategies $\boldsymbol{\phi}$, but not convex in $\boldsymbol{\phi}$.
Geodesic convexity is a generalization of convexity to Riemannian manifolds, 
or more simply (in our context), convexity subject to a variable transformation.
We refer the readers to Riemannian optimization textbooks, e.g., \cite{boumal2020introduction}, for formal definition and optimization techniques of geodesic convex functions.
Suppose request input rate vector $\boldsymbol{r} = [r_i(k)]_{i\in\mathcal{V}, k \in \mathcal{C}}$ has every element strictly greater than $0$, it is easy to see that the set of feasible routing strategies $\mathcal{D}_{\boldsymbol{\phi}}$ and the set of feasible flows $\mathcal{D}_{\boldsymbol{f}}$ defined as
\begin{equation*}
\begin{aligned}
    \mathcal{D}_{\boldsymbol{f}} = \left\{[f_{ij}(k)]_{(i,j)\in\mathcal{E}, k\in\mathcal{C}}\bigg| \sum_{j\in\mathcal{O}(i)}f_{ij}(k) = \sum_{j\in\mathcal{O}(i)}f_{ji}(k) + r_i(k) \right\}
\end{aligned}
\end{equation*}
has a one-to-one mapping.
Therefore, there exists a unique variable transformation $\boldsymbol{\phi}$ to $\boldsymbol{f}$, so that the routing cost $\sum_{(i,j) \in \mathcal{E}} D_{ij}(F_{ij})$ is a convex function of $\boldsymbol{f}$ (this is obvious due to the convex assumption of function $D_{ij}$).
Therefore, the routing cost $\sum_{(i,j) \in \mathcal{E}} D_{ij}(F_{ij})$ is a geodesic convex function of $\boldsymbol{\phi}$, which implies $T(\boldsymbol{\phi},\boldsymbol{y})$ is a summation of a convex function and a geodesic convex function.
To the best of our knowledge, we are the first to formulate and tackle such problem with a provable bound. 

\section{Proof of Corollary \ref{cor_cache_existance}}
\label{proof_cor_existance}

To prove the existence of a global optimal solution that satisfies condition \eqref{condition_BoundedGap_cache}, without loss of generality, we assume the global optimal solution to \eqref{Objective_cache} exist, and with finite objective value. 
In this proof, we show that for any given $(\boldsymbol{\phi}^*,\boldsymbol{y}^*)$ that optimally solves \eqref{Objective_cache} with finite objective value $T(\boldsymbol{\phi}^*,\boldsymbol{y}^*)$, there always exists a $(\boldsymbol{\phi},\boldsymbol{y})$ that satisfies \eqref{condition_BoundedGap_cache} and with $T(\boldsymbol{\phi},\boldsymbol{y}) = T(\boldsymbol{\phi}^*,\boldsymbol{y}^*)$. That is, we construct such $(\boldsymbol{\phi},\boldsymbol{y})$ from $(\boldsymbol{\phi}^*,\boldsymbol{y}^*)$.

We first observe that the caching part of condition \eqref{condition_BoundedGap_cache} must be satisfied with $(\boldsymbol{\phi}^*,\boldsymbol{y}^*)$, namely, to show for all $i$ and $k$ it holds that
\begin{equation}
B^\prime_{i}(Y_{i}^*) \begin{cases}
= t_{i}^*(k) \delta_{i}^*(k), \quad \text{if } y_{i}^*(k) > 0,
\\ \geq t_{i}^*(k) \delta_{i}^*(k), \quad \text{if } y_{i}^*(k) = 0,
\end{cases}
\label{Proof_cache_OptExistence_1}
\end{equation}
where $\delta_{i}^*(k)$ is defined coherently as in \eqref{condition_BoundedGap_cache}:
\begin{equation*}
    \delta_{i}^*(k) = \begin{cases} \min\left(\min\limits_{j^\prime \in \mathcal{N}(i)}\left(D^\prime_{j^\prime i}\left(F_{j^\prime i}^*\right) + \frac{\partial T^*}{\partial r_{j^\prime}(k)}\right),
     \frac{B^\prime_{i^\prime}(Y_{i^\prime}^*)}{t_{i}^*(k)}\right), 
  \,\text{if } t_{i}^*(k) > 0
    \\ \min\limits_{j^\prime \in \mathcal{N}(i)}\left(D^\prime_{j^\prime i}\left(F_{j^\prime i}^*\right) + \frac{\partial T^*}{\partial r_{j^\prime}(k)}\right), \quad  \text{if } t_{i}^*(k) = 0
    \end{cases}
\end{equation*}

In fact, \eqref{Proof_cache_OptExistence_1} holds trivially from the KKT condition in Theorem \ref{thm_necessary_cache} applied to $(\boldsymbol{\phi}^*,\boldsymbol{y}^*)$. Specifically, for any $i$,$k$, if $t_i^*(k) >0$, then \eqref{Proof_cache_OptExistence_1} holds as it is equivalent to \eqref{condition_necessary_cache}. If $t_i^*(k) = 0$, then we must have $y_{i}^*(k) = 0$, as caching $k$ at $i$ will provide no improvement to the link cost, but only increase to the caching cost.
Therefore \eqref{Proof_cache_OptExistence_1} holds regardless of the actual value of $\delta_{i}^*(k)$.
Hence, we construct the caching variable as 
\begin{equation*}
    \boldsymbol{y} = \boldsymbol{y}^*.
\end{equation*}
We next construct routing variable $\boldsymbol{\phi}$. 
Let set $\mathcal{C}^* = \left\{(i,k) \big| t_i^*(k) \neq 0\right\}$.
We construct the elements of $\boldsymbol{\phi}$ corresponding to $\mathcal{C}^*$ be identical to those of $\boldsymbol{\phi}^*$, i.e.,
\begin{equation*}
    \phi_{ij}(k) = \phi_{ij}^*(k), \quad \forall (i,k) \in \mathcal{C}^*, \, \forall j \in \mathcal{N}(i).
\end{equation*}
Recall that when $\boldsymbol{y} = \boldsymbol{y}^*$ and $\phi_{ij}(k) = \phi_{ij}^*(k)$ for $(i,k) \in \mathcal{C}^*$, and by \eqref{FlowConservation_cache}, we know that no mater what $\phi_{ij}(k)$ we assign for $(i,k) \not\in \mathcal{C}^*$ (i.e., those $i$,$k$ with $t_{i}^*(k) = 0$), it always holds that $f_{ij}(k) = f_{ij}^*(k)$ for all $i$,$j$,$k$, and thus $t_{i}(k) = t_i^*(k)$ for all $i$,$k$.
Therefore, no mater what $\phi_{ij}(k)$ we assign for $(i,k) \not\in \mathcal{C}^*$, let $\mathcal{C} = \left\{(i,k) \big| t_i(k) \neq 0\right\}$, it holds that $\mathcal{C} = \mathcal{C}^*$ and $T(\boldsymbol{\phi},\boldsymbol{y}) = T(\boldsymbol{\phi}^*,\boldsymbol{y}^*)$, i.e., $(\boldsymbol{\phi},\boldsymbol{y})$ is also a global optimal solution to \eqref{Objective_cache}.

We further construct $\phi_{ij}(k)$ for $(i,k) \not\in \mathcal{C}$. 
For each $(i,k) \not\in \mathcal{C}$, we simply pick one $j \in \mathcal{N}(i)$ with
\begin{equation}
    j \in \arg\min_{j^\prime \in \mathcal{N}(i)} \left( D^\prime(F_{j^\prime i}^*) + \frac{\partial T^*}{\partial r_{j^\prime}(k)}\right)
\label{Proof_cache_OptExistence_2}
\end{equation}
and let $\phi_{ij}(k) = 1$, while let $\phi_{ij^\prime}(k) = 0$ for all other $j^\prime \neq j$.
Note that with out loss of generality, we assume the optimal solution $\boldsymbol{\phi}^*$ is loop-free.
Thus such construction is always feasible (i.e., there always exists a set of $\phi_{ij}(k)$ satisfying \eqref{Proof_cache_OptExistence_2} for all $(i,k) \not\in \mathcal{C}$), for example, the construction could start at sinks (nodes with $y_i(k) = 1$) and destinations (nodes $i \in \mathcal{S}_k$), and propagates in the upstream order.

Therefore till this point, we have constructed $(\boldsymbol{\phi},\boldsymbol{y})$ by specifying all elements of it. 
Since $(\boldsymbol{\phi},\boldsymbol{y})$ is also a global optimal solution to \eqref{Objective_cache}, by Theorem \ref{thm_necessary_cache}, we know \eqref{condition_necessary_cache} holds for $(\boldsymbol{\phi},\boldsymbol{y})$.
Thus \eqref{condition_BoundedGap_cache} hold for $(i,k) \in \mathcal{C}$, simply by dividing $t_i(k)$ from both side of \eqref{condition_necessary_cache}.
For $(i,k) \not\in \mathcal{C}$, we know the routing part of \eqref{condition_BoundedGap_cache} also holds, due to the method \eqref{Proof_cache_OptExistence_2} we constructed those $\phi_{ij}(k)$.
The caching part of \eqref{condition_BoundedGap_cache} holds as well to $(i,k) \not\in \mathcal{C}$, since we must have $y_{i}(k) = y_{i}^*(k) = 0$.

Consequently, the constructed $(\boldsymbol{\phi},\boldsymbol{y})$ is a global optimal solution to \eqref{Objective_cache} and satisfies \eqref{condition_BoundedGap_cache}, which completes the proof.

\section{Proof of Corollary \ref{cor_cache_2}}
\label{proof_cor_f_increase}
We first prove the case $\boldsymbol{\phi}^\dagger \geq \boldsymbol{\phi}$. 
Let $p_{ik} = 1 - y_{i}(k) = \sum_{j \in \mathcal{V}}\phi_{ij}(k)$.
In this case, it is obvious that 
\begin{equation*}
    p_{ik}^\dagger \geq p_{ik}, \quad \forall i \in \mathcal{V}, k \in \mathcal{C}.
\end{equation*}
Meanwhile, since $\boldsymbol{\phi}^\dagger \geq \boldsymbol{\phi}$, given the exogenous input rates $r_i(k)$ unchanged, the corresponding link flows are also non-decreasing. Namely, 
\begin{equation*}
\begin{aligned}
    f_{ij}^\dagger(k) &\geq f_{ij}(k), \quad \forall (i,j) \in \mathcal{E}, k\in \mathcal{C},
    \\t_i^\dagger(k) &\geq t_i(k), \quad \forall i \in \mathcal{V}, k\in \mathcal{C}.
\end{aligned}
\end{equation*}
Combing the above with Theorem \ref{thm_BoundedGap_cache} and noticing $\delta_{i}(k) \geq 0$, we have $T(\boldsymbol{\phi}^\dagger,\boldsymbol{y}^\dagger) \geq T(\boldsymbol{\phi},\boldsymbol{y})$. Same reasoning applies for case $\boldsymbol{\phi}^\dagger \leq \boldsymbol{\phi}$, which completes the proof.

\section{Loops are always suboptimal with integer caching decisions}
\label{proof_integer_no_loop}
\begin{prop}
Let $(\boldsymbol{\phi},\boldsymbol{x})$ (where $\boldsymbol{x})$ is binary caching decisions) be feasible to \eqref{Objective_cache} and contains a routing loop $(l_1, l_2, \cdots, l_{|l|})$ with strictly positive flow, then there always exist another feasible $(\boldsymbol{\phi}^\prime,\boldsymbol{x}^\prime)$ such that $T(\boldsymbol{\phi}^\prime,\boldsymbol{x}^\prime) < T(\boldsymbol{\phi},\boldsymbol{x})$
\end{prop}
\begin{proof}
For simplicity, we only prove the case $|l| = 3$, i.e., the loop of $2$ nodes $i$ and $j$. Similar reasoning could be applied to loop of arbitrary length.

We assume $|\mathcal{C}| = 1$ and omit the notation of item.
Without loss of generality, we assume $\mathcal{N}(i) = \{j , p\}$ and $\mathcal{N}(i) = \{i , q\}$, that is, we abstract all neighbors of $i$ other than $j$ to a single node $p$, and abstract all neighbors of $j$ other than $i$ to a single node $q$.
Moreover, we let $\boldsymbol{x}^\prime = \boldsymbol{x}$, and $\boldsymbol{\phi}_{v}^\prime = \boldsymbol{\phi}_{v}$ for all $v \neq i,j$.
We assume all input requests to node $i$ that are not forwarded by $j$ have steady rate $r_i$ (this could contain the original $r_i$ and all endogenous input from other nodes than $j$), and all input requests to node $j$ that are not forwarded by $i$ have steady rate $r_j$.
We demonstrate in Fig. \ref{fig_noloop} this simplified network.

\begin{figure}[htbp]
\centerline{\includegraphics[width=1\linewidth]{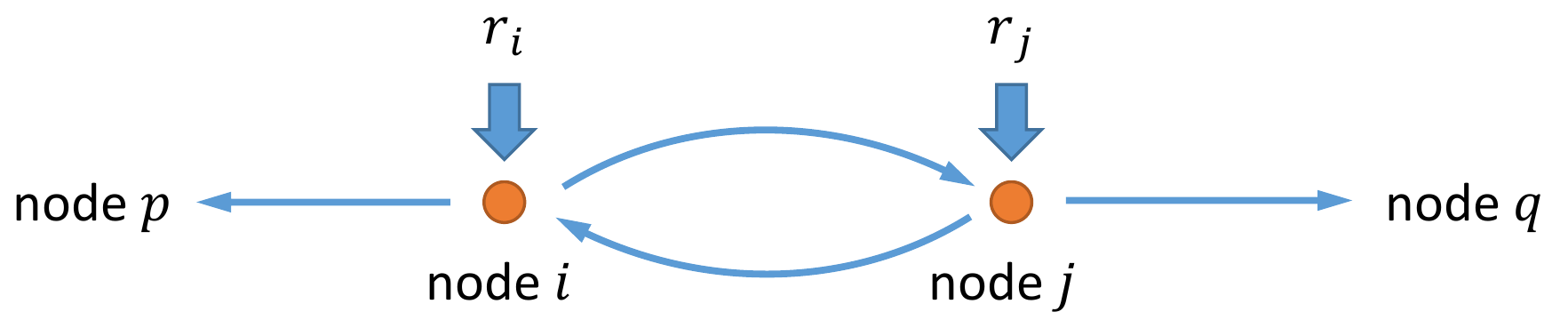}}
\caption{Simplified example with loop $i \to j \to i$}
\label{fig_noloop}
\end{figure}

Since the loop flow is strictly positive, we know $x_i = x_j = 0$.
Moreover, 
\begin{equation*}
    t_i = r_i + \phi_{ji} t_j  = r_i + \phi_{ji}(r_j + \phi_{ij}t_i),
\end{equation*}
thus 
\begin{equation*}
    t_i = \frac{r_i + \phi_{ji}r_j}{1 - \phi_{ij} \phi_{ji} },
\end{equation*}
and similarly
\begin{equation*}
    t_j = \frac{r_j + \phi_{ij}r_i}{1 - \phi_{ij} \phi_{ji} }.
\end{equation*}

Viewing node $i$ and $j$ as a group, we look at the out-going flows from them to the rest of the network. Specifically,
\begin{equation*}
    f_{ip} = t_i \phi_{ip} = t_i (1 - \phi_{ij}) = \frac{1 - \phi_{ij}}{1 - \phi_{ij} \phi_{ji} }\left(r_i + \phi_{ji}r_j\right),
\end{equation*}
and similarly
\begin{equation*}
    f_{jq} = \frac{1 - \phi_{ji}}{1 - \phi_{ij} \phi_{ji} }\left(r_j + \phi_{ij}r_i\right).
\end{equation*}
Thus we have 
\begin{equation*}
    f_{ip} + f_{jq} = \frac{(1 - \phi_{ij})(r_i + \phi_{ji}r_j) + (1 - \phi_{ji})(r_j + \phi_{ij}r_i)}{(1 - \phi_{ij} \phi_{ji})}
     = r_i + r_j,
\end{equation*}
namely, the flow rate from node $i$ and $j$ to the rest of the network always equals their input flow rate, regardless of the loop variables $\phi_{ij}$ and $\phi_{ji}$.

Therefore, it must holds that either $f_{ip} \geq r_i$ or $f_{jq} \geq r_j$. Without loss of generality, we assume $f_{jq} \geq r_j$ ,then $f_{ip} \leq r_i$.  
We then let the new variables $\boldsymbol{\phi}_i^\prime$ and $\boldsymbol{\phi}_j^\prime$ be
\begin{equation*}
    \phi_{ij}^\prime = \frac{f_{ip}}{r_i}, \quad \phi_{ip}^\prime = 1 - \frac{f_{ip}}{r_i},
\end{equation*}
and 
\begin{equation*}
    \phi_{ji}^\prime = 0, \quad \phi_{jq}^\prime = 1,
\end{equation*}
which implies that 
\begin{equation*}
    f_{ip}^\prime = f_{ip}, \quad f_{jq}^\prime = f_{jq}.
\end{equation*}

Therefore, with such constructed $(\boldsymbol{\phi}^\prime,\boldsymbol{x}^\prime)$, the loop is eliminated, and all other flows in the network remains unchanged because $f_{ip} = f_{ip}^\prime$ and $f_{jq} = f^\prime_{jq}$.
Moreover, within the two nodes $i$ and $j$, it is obvious that $f_{ij}^\prime < f_{ij}$ and $f_{ji}^\prime < f_{ji}$.
Thus the total cost must strictly decrease from $(\boldsymbol{\phi},\boldsymbol{x})$ to  $(\boldsymbol{\phi}^\prime,\boldsymbol{x}^\prime)$ since the loop flow strictly decreases and functions $D_{ij}(\cdot)$ and $D_{ji}(\cdot)$ are monitonically increasing, which completes the proof.
\end{proof}

\section{Routing Loops in optimal solution}
\label{routing_loops}
\begin{figure}[htbp]
\centerline{\includegraphics[width=1\linewidth]{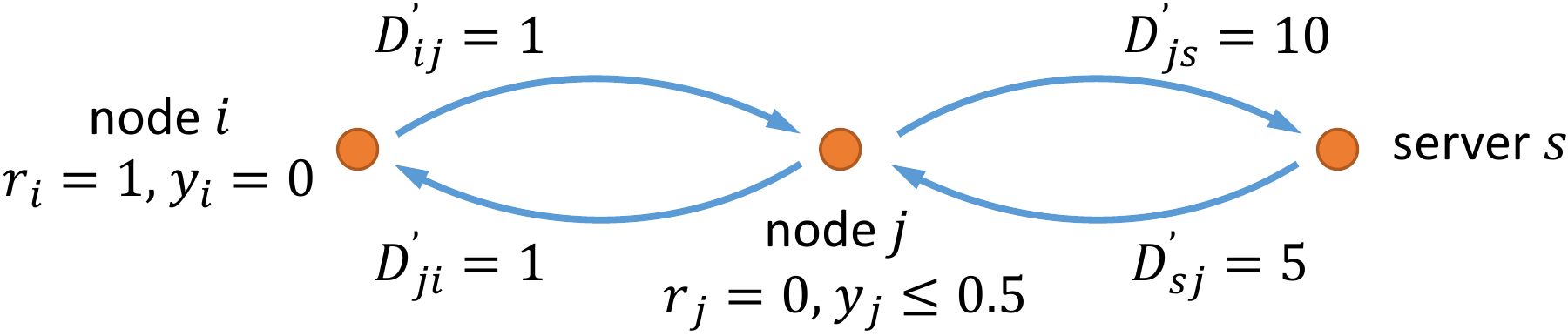}}
\caption{ Toy example for the existence of loops in the optimal solution to \eqref{Objective_cache}. There is only one item requested by $i$, and $s$ is the designated server. Only node $j$ has the capability to cache and $y_j$ is upper bounded by $0.5$. Link costs are all linear, their marginals $D^\prime$ are marked on the figure.}
\label{fig_loop}
\end{figure}

An example is given in Figure \ref{fig_loop}:
If loops are not allowed, the best solution is apparently $\phi_{ij} = 1$, $\phi_{js} = 0.5$ and $y_j = 0.5$, with $F_{ij} = F_{js} = 0$, $F_{ji} = 1$, $F_{sj} = 0.5$, and total cost $T = 3.5$.
However, if loops are allowed, with the solution $\phi_{ij} = 1$, $\phi_{ji} = 0.5$ and $y_j = 0.5$, we have $F_{ij} = 1$, $F_{ji} = 2$, $F_{js} = F_{sj} = 0$, and total cost $T = 3$.
One can verify that the latter solution, with a loop $i \to j \to i$, satisfies condition \eqref{condition_BoundedGap_cache} and is in fact the optimal solution to \eqref{Objective_cache}.
Nevertheless, such optimal solutions with loops exist only for the continuous problem \eqref{Objective_cache} in a mathematical sense.
For the integer situation, we show in {\rm Appendix \ref{proof_integer_no_loop}} that loops always yield suboptimal solutions.  

\section{Dynamic blocked node sets}
\label{dynamic_blocked_set}
In \cite{gallager1977minimum,xi2008node,wiopt22}, set $\mathcal{B}_i^t(k)$ is defined as the nodes $j$ such that either $(i,j) \not\in \mathcal{E}$ or $\partial T/\partial r_j(k) > \partial T/\partial r_i(k)$ at $t$-th period, because in their case, $\partial T/\partial r_i(k)$ must be monotonically decreasing on any request path for $k$ in the optimal solution.
However, such definition is not applicable to us, since the decreasing order may not hold (e.g., in Figure \ref{fig_loop}). To this end, besides other network-embedded loop prevention methods, we next offer a new definition of $\mathcal{B}_i^t(k)$ valid for our case.

\begin{figure}[htbp]
\centerline{\includegraphics[width=1\linewidth]{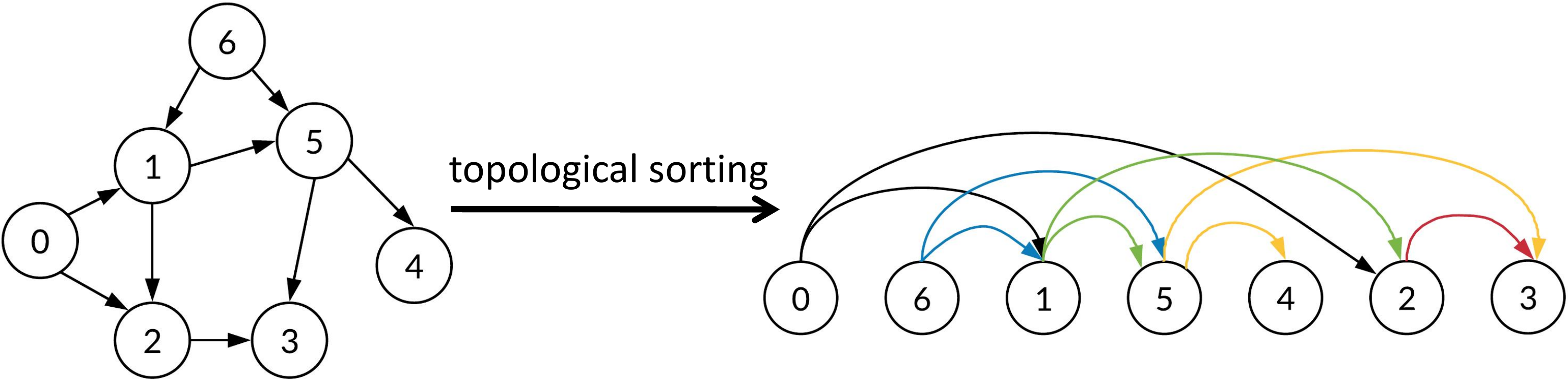}}
\caption{ Example of topological sorting on a DAG. The result forms a total order (the left-to-right order of $0,6,1,5,4,2,3$) on the set of all nodes.}
\label{fig_toposort}
\end{figure}

Suppose $\boldsymbol{\phi}^t$ is loop-free. 
Then for any $k$, the subgraph of $\mathcal{G}$ consisting of $\mathcal{V}$ and $(i,j) \in \mathcal{E} : \phi_{ij}^t(k) > 0$ forms a directed acyclic graph (DAG). 
A total order on $\mathcal{V}$ complying with this DAG must exist and can be calculated by a topological sorting\footnote{Topological sorting can be accomplished in a distributed manner, see, e.g., \cite{ma1997efficient}. The implementation detail of topological sorting is beyond the scope of this paper.}, illustrated in Figure \ref{fig_toposort}.

We use notation $ i >^{(k)} j$ if node $i$ is in front of node $j$ in the topological-sorted total order for item $k$, e.g., we have $ i >^{(k)} l$ if $\phi_{ij}(k) > 0$ and $\phi_{jl}(k) > 0$.
Then we construct the set $\mathcal{B}_i^t(k)$ as follows,
\begin{equation}
    \mathcal{B}_i^{t+1}(k) = \left\{ j \in \mathcal{V} \big| (i,j) \not\in \mathcal{E} \text{ or } j >^{(k)} i \text{ in $t$-th period} \right\}, \quad \forall t \geq 0.
\label{alg_block_set}
\end{equation}

\begin{lem}
\label{Lemma_loopfree}
If the initial routing scheme $\boldsymbol{\phi}^0$ is loop-free, and the subsequent variables are generated by \eqref{variable_update} with the blocked node set $\mathcal{B}_i^t(k)$ defined as \eqref{alg_block_set}, then $\boldsymbol{\phi}^t$ is loop-free for all $t \geq 1$.
\end{lem}
\begin{proof}
We assume that at $t$-th period the routing scheme $\boldsymbol{\phi}^t$ is loop-free, and the blocked node sets are practiced when updating to $\boldsymbol{\phi}^{t+1}$.
If there exist is a loop in $\boldsymbol{\phi}^{t+1}$ for item $k$, there must exist $i,j \in \mathcal{V}$ such that there exists a request path from $i$ to $j$ and vice versa. 
However, according to the total order ``$>^{(k)}$'' in $t$-th period which generates the blocked node sets $\mathcal{B}_i^{t+1}(k)$, either $i >^{(k)} j$ or $j >^{(k)} i$ must hold.
Without loss of generality, we assume $i >^{(k)} j$.
Therefore, there must exist an edge $(p,q) \in \mathcal{E}$ on the path from $j$ to $i$ w.r.t. item $k$, such that $q >^{(k)} p$ (otherwise we must have $j >^{(k)} i$ by the transitivity of a total order).
However in this case, the existence of such $(p,q)$ with $\phi_{pq}^{t+1}(k) > 0$ violates the rule of blocked node sets, since $q>^{(k)} p$ implies that $q \in \mathcal{B}_p^{t+1}(k)$.
Consequently, the new routing scheme $\boldsymbol{\phi}^{t+1}$ must contain no loops, which completes the proof. 
\end{proof}

\section{Proof of Theorem \ref{thm_convergence}}
\label{proof_thm_convergence}
Our proof of asynchronous convergence prove follows \cite{xi2008node}.
We first show that, when all other node but node $i$ keep their variable unchanged (which we denote by $(\boldsymbol{\phi}_{\text{-}i},\boldsymbol{y}_{\text{-}i})$), the total cost $T$ is convex in $(\boldsymbol{\phi}_i, \boldsymbol{y}_i)$. 

\begin{lem}
For $i \in \mathcal{V}$, total cost $T$ is jointly convex in $(\boldsymbol{\phi}_i, \boldsymbol{y}_i)$, provided that $(\boldsymbol{\phi}_{\text{-}i},\boldsymbol{y}_{\text{-}i})$ is fixed.
\label{lemma_convex_single_node}
\end{lem}
\begin{proof}
Provided that the cost functions $B_i(\cdot)$ and $D_{ij}(\cdot)$ are convex, it is sufficient to show that when $(\boldsymbol{\phi}_{\text{-}v}, \boldsymbol{y}_{\text{-}v})$ is fixed,  $Y_i$ and  $F_{ij}$ are linear in $(\boldsymbol{\phi}_v, \boldsymbol{y}_v)$, for any $i \in \mathcal{V}$ and  $(i,j) \in \mathcal{E}$.

First, since $Y_i$ is not relevant to $(\boldsymbol{\phi}_v, \boldsymbol{y}_v)$ for any $i \neq v$, and $Y_v = \sum_{k \in \mathcal{C}} y_v(k)$, we know every cache size $Y_i$ is linear in $(\boldsymbol{\phi}_v, \boldsymbol{y}_v)$.

Next, we write the link flow $F_{ij}$ in a multi-linear form of $(\boldsymbol{\phi}, \boldsymbol{y})$.
As an extension of $p_{vk}$ defined in Section \ref{sec:fixed routing}, let $\mathcal{P}_{vi}(k)$ be the set of paths from $v$ to $i$ corresponding to item $k$.
That is, $\mathcal{P}_{vi}(k)$ is the set of all paths $p$ such that $p = \left(p^1,p^2,\cdots,p^{|p|}\right)$, where $p^1 = v$, $p^{|p|} = i$ and $\phi_{p^l p^{l+1}}(k) > 0$ for all $l = 1,\cdots,|p|-1$.
Then the following holds,
\begin{equation*}
    t_i(k) = \sum_{v \in \mathcal{V}} r_v(k) \sum_{p \in \mathcal{P}_{vi}(k)} \prod_{l = 1}^{|p|-1} \phi_{p^l p^{l+1}}(k)
\end{equation*}
and thus
\begin{equation}
\begin{aligned}
    &F_{ji} = \sum_{k \in \mathcal{C}} t_i(k)\phi_{ij}(k) 
    \\&=  \sum_{k \in \mathcal{C}} \phi_{ij}(k) \left( \sum_{v \in \mathcal{V}} r_v(k) \sum_{p \in \mathcal{P}_{vi}(k)} \prod_{l = 1}^{|p|-1} \phi_{p^l p^{l+1}}(k)\right)
\end{aligned}
    \label{proof_convex_Fji}
\end{equation}
Recall that we assumed no loops are formed, then \eqref{proof_convex_Fji} is a multi-linear function of $\boldsymbol{\phi}$, thus $F_{ji}$ is linear in $(\boldsymbol{\phi}_v, \boldsymbol{y}_v)$ provided $(\boldsymbol{\phi}_{\text{-}v}, \boldsymbol{y}_{\text{-}v})$ is fixed, which completes the proof.
\end{proof}

Without loss of generality, for any period $t$, we assume $t_{v(t)}(k) > 0$ for some $k$ (otherwise, node $v(k)$ is not processing any request, we can skip this period in the sequence).
Thus for node $v(t)$ and that item $k$, the condition \eqref{condition_BoundedGap_cache} in the feasible set and with the restriction of fixed blocked nodes is essentially a KKT condition for a convex problem.
Therefore, it is a necessary and sufficient condition for the optimality of the subproblem of node $v(k)$.
Then, if $T^t$ is not the optimal solution of this subproblem, that is, if the condition \eqref{condition_BoundedGap_cache} at $v(t)$ with $\mathcal{N}(v(t))$ replaced by $\mathcal{N}(v(t)) \backslash \mathcal{B}_{v(t)}(k)$ is not satisfied by $(\boldsymbol{\phi}_{v(t)}, \boldsymbol{y}_{v(t)})$,
it holds that gradient projection with sufficiently small stepsize must yield a decreasing total cost (for a detailed stepsize value, please refer to 
\cite{gallager1977minimum,bertsekas1997nonlinear}).
On the other hand, if condition \eqref{condition_BoundedGap_cache} at $v(t)$ is satisfied, the current (optimal solution) to this subproblem is kept unchanged by Algorithm \ref{alg_GP}, and we can also skip this period in the sequence.

Consequently, ruling out the cases of (1) node $v(t)$ has $t_{v(t)}(k) = 0$ for all $k$, and (2) strategy $(\boldsymbol{\phi}_{v(t)}, \boldsymbol{y}_{v(t)})$ is already optimal to the subproblem, Algorithm \ref{alg_GP} must yield a strictly decreasing objective sequence, i.e., $T^{t+1} < T^t$.
(Note that the assumption $\lim_{t \to \infty}\left|\mathcal{T}_i\right| = \infty$ implies that the system state will not be stuck at these two special cases, as long as there are other nodes not satisfying \eqref{condition_BoundedGap_cache}.)

Moreover, since $T^0 < \infty$, it is obvious that the feasible set with the restriction of blocked nodes is a compact set.
Therefore, the sequence $\left\{(\boldsymbol{\phi}^t, \boldsymbol{y}^t)\right\}_{t=0}^{\infty}$, which generates a strictly decreasing objective in a compact set, must have a subsequence converging to a limit point $(\boldsymbol{\phi}, \boldsymbol{y})$, which can not be further updated by \eqref{variable_update}.
Namely, the limit point $(\boldsymbol{\phi}, \boldsymbol{y})$ must satisfies condition \eqref{condition_BoundedGap_cache} with the restriction of blocked node sets, i.e., with $\mathcal{N}(i)$ replaced by $\mathcal{N}(i) \backslash \mathcal{B}_i(k)$.

We remark that the condition of sufficiently small stepsize is adopted as it guarantees the convergence of gradient projection. Nevertheless, in practical implementation, one may choose larger stepsize to speed up the convergence.
A number of Newton-like convergence acceleration methods can be used to improve our vanilla version gradient projection, e.g., by a second order method by \cite{bertsekas1984second} or by further upper bounding the Hessian-inverse by \cite{xi2008node}. 
We leave the implementation of these advanced techniques as a further direction.

\section{Distributed randomized rounding}
\label{appendix_rounding}
Our method is a generalization of the distributed rounding technique devised in \cite{ioannidis2018adaptive}.
Specifically, for each period, \texttt{DRR} generates a univariate mapping $P$, which maps an input $t \in [0,1]$ to a set of items $P(t) \subseteq \mathcal{C}$ to be cached, based on a continuous cache strategy $ \boldsymbol{y} = [y(k)]_{k \in \mathcal{C}}$.
The mapping $P$ is generated from a bar-coloring procedure with the following rules:
(1) Each item $k$ is assigned with a distinct color $c_k$.
(2) Start from the first line with item $k = 1$, draw a horizontal bar of length $y(1)$ (the height of the bar is unimportant), and color it with $c_1$.
(3) After drawing and coloring the bar for item $k$, draw the bar for item $k+1$ of length $y(k+1)$ starting from the end of the previous bar, and color it with $c_{k+1}$. 
(4) When the length of current line reaches $1$, start over at a new line.
(5) End when bars for all $|\mathcal{C}|$ items are finished.

\begin{figure}[]
\centerline{\includegraphics[width=0.9\linewidth]{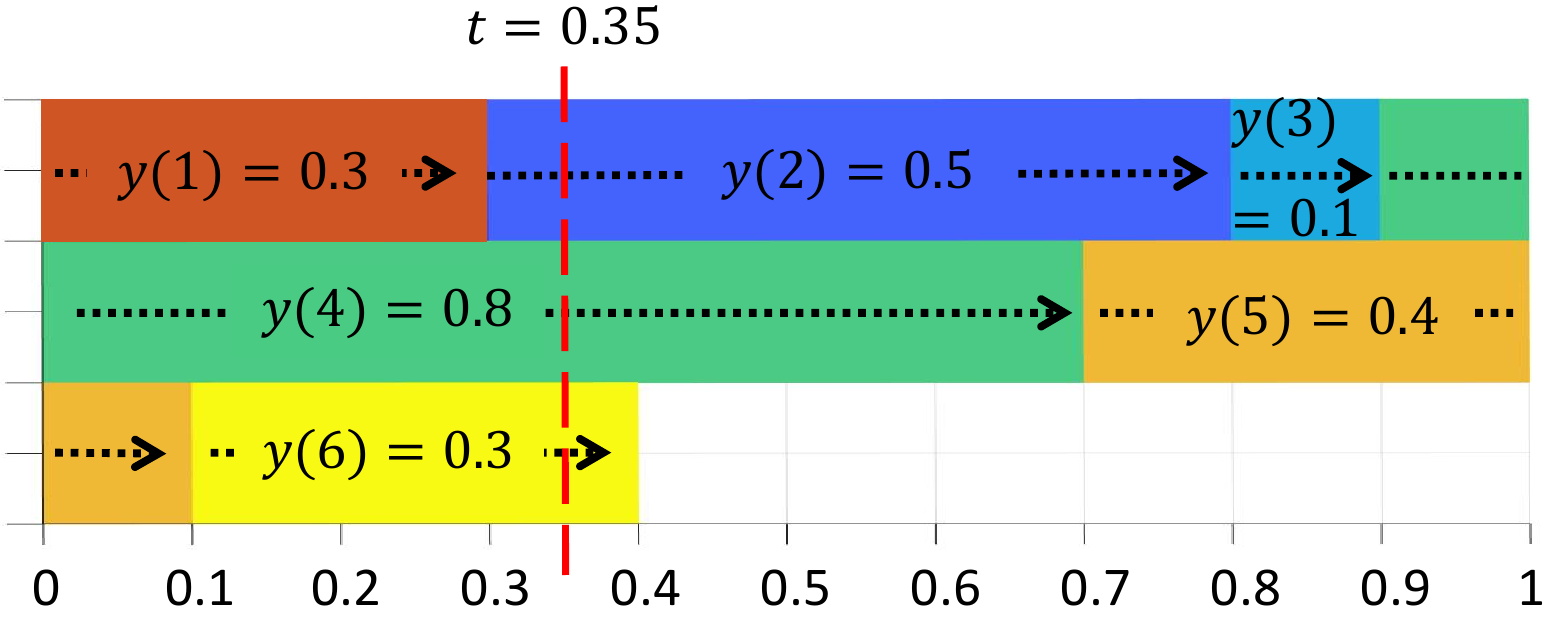}}
\caption{Graphical illustration of bar-coloring with $|\mathcal{C}| = 6$ and $\boldsymbol{y} = (0.3, 0.5, 0.1, 0.8, 0.4, 0.3)$ for the mapping $P$ in \texttt{DRR}. The bars of item $2$, $4$ and $6$ crosses position $t = 0.35$, thus $P(0.35) = \left\{2,4,6\right\}$. }
\label{fig_colormap}
\end{figure}

Then $P(t), t\in[0,1]$ is the set of $k\in\mathcal{C}$ such that the bar with color $c_k$ goes across the horizontal position $t$. 
Figure \ref{fig_colormap} graphically demonstrates the construction of mapping $P$.

We denote by $P_i^t$ the mapping generated by \texttt{DRR} at node $i$ for $t$-th period. Then for the $m$-th slot of $t$-th period, node $i$ picks a u.a.r variable $t_i^{t,m} \in [0,1]$ independent of other nodes and other slots, and makes its caching decision by $\boldsymbol{x}_i^{t,m} = P_i^t(t_i^{t,m})$. 
With such rounding method, the actual cache sizes $X_i$ are kept relatively stable at all nodes around their $Y_i$.

Note that \texttt{DRR} does not guarantee the independence of $x_{i}(k)$ among different $k$. This is a tradeoff that has to be made for steady cache sizes. 

\end{document}